%% file: main.tex
\documentclass[sigconf]{acmart}
\AtBeginDocument{%
  \providecommand\BibTeX{{%
    \normalfont B\kern-0.5em{\scshape i\kern-0.25em b}\kern-0.8em\TeX}}}

\usepackage{hyperref}       
\usepackage{url}            
\usepackage{booktabs}       
\usepackage{amsfonts}       
\usepackage{nicefrac}       
\usepackage{microtype}      
\usepackage{lipsum}		
\usepackage{doi}
\usepackage{subcaption}
\usepackage{colortbl}
\usepackage{balance}

\usepackage[strict]{changepage}

\usepackage{framed}
\definecolor{formalshade}{rgb}{0.85,1,0.85}
\definecolor{darkblue}{rgb}{0.0,0.6,0.30}
\newenvironment{formal}{%
  \MakeFramed{\advance\hsize-\width\FrameRestore}%
  \noindent\hspace{-4.55pt}
  \begin{adjustwidth}{}{7pt}%
}
{%
  \end{adjustwidth}\endMakeFramed%
}

\setcopyright{none}
\renewcommand\footnotetextcopyrightpermission[1]{} 

\begin{document}

\title[Your Attack Is Too DUMB]{Your Attack Is Too DUMB: \\Formalizing Attacker Scenarios for Adversarial Transferability}


\author{Marco Alecci}
\orcid{0000-0002-5963-4599}
\affiliation{%
  \institution{SnT, University Of Luxembourg}
  \streetaddress{6, rue Richard Coudenhove-Kalergi}
  \city{Luxembourg}
  \country{Luxembourg}}
\email{marco.alecci@uni.lu}

\author{Mauro Conti}
\orcid{0000-0002-3612-1934}
\affiliation{%
  \institution{University of Padova}
  \streetaddress{Via Trieste, 63}
  \city{Padua}
  \country{Italy}}
\email{mauro.conti@unipd.it}

\author{Francesco Marchiori}
\orcid{0000-0001-5282-0965}
\affiliation{%
  \institution{University of Padova}
  \streetaddress{Via Trieste, 63}
  \city{Padua}
  \country{Italy}}
\email{francesco.marchiori.4@phd.unipd.it}

\author{Luca Martinelli}
\affiliation{%
  \institution{University of Padua}
  \city{Padua}
  \country{Italy}}
\email{luca.martinelli.1@studenti.unipd.it}

\author{Luca Pajola}
\authornote{Corresponding author.}
\orcid{0000-0002-6749-6608}
\affiliation{%
  \institution{University of Padua}
  \city{Padua}
  \country{Italy}}
\email{luca.pajola@phd.unipd.it}


\begin{abstract}
Evasion attacks are a threat to machine learning models, where adversaries attempt to affect classifiers by injecting malicious samples. 
An alarming side-effect of evasion attacks is their ability to transfer among different models: this property is called \textit{transferability}. Therefore, an attacker can produce adversarial samples on a custom model (surrogate) to conduct the attack on a victim's organization later. 
Although literature widely discusses how adversaries can transfer their attacks, their experimental settings are limited and far from reality. 
For instance, many experiments consider both attacker and defender sharing the same dataset, balance level (i.e., how the ground truth is distributed), and model architecture. 
\par
In this work, we propose the \textbf{DUMB} attacker model. This framework allows analyzing if evasion attacks fail to transfer when the training conditions of surrogate and victim models differ. DUMB considers the following conditions: \textbf{D}ataset so\textbf{U}rces, \textbf{M}odel architecture, and the \textbf{B}alance of the ground truth. 
We then propose a novel testbed to evaluate many state-of-the-art evasion attacks with DUMB; the testbed consists of three computer vision tasks with two distinct datasets each, four types of balance levels, and three model architectures. 
Our analysis, which generated 13K tests over 14 distinct attacks, led to numerous novel findings in the scope of transferable attacks with surrogate models. In particular, mismatches between attackers and victims in terms of dataset source, balance levels, and model architecture lead to non-negligible loss of attack performance. 
\end{abstract}





\maketitle
\pagestyle{plain}

\input{Sections/01-Introduction.tex}
\input{Sections/02-Preliminaries.tex}
\input{Sections/03-Methodology.tex}
\input{Sections/04-ExperimentalSettings.tex}
\input{Sections/05-Results.tex}
\input{Sections/06-Conclusion.tex}


\balance
\bibliographystyle{ACM-Reference-Format}
\bibliography{references}

\clearpage
\appendix

\input{Sections/Appendix/A1-Attack_Examples}

\end{document}

%% file: Sections/01-Introduction.tex
\section{Introduction}\label{sec.introduction}
\textit{Evasion attacks} consist in crafting a sample to produce a misclassification in a target Machine Learning (ML) model. 
With the integration of ML models in deployed real-life systems, the cybersecurity community increased its interest in studying how \textit{attackers} can exploit ML vulnerabilities for some advantages. 
For instance, an attacker might try to make a hateful sentence look non-hateful~\cite{grondahl2018all} or botnets legitimate applications~\cite{apruzzese2018evading}.
\par
Although effective in theory, conducting evasion attacks in real-world scenarios is challenging since malicious actors cannot access target models' information (e.g., the gradient)~\cite{apruzzese2022real}. 
Adversarial samples \textit{transferability} is a possible solution investigated in prior works~\cite{papernot2016transferability}: the attacker feeds the victim's model with adversarial samples computed by leveraging an own \textit{surrogate model}. 
\par
While attempting to test the robustness of ML models of top IT companies (through their official APIs), we realized that not even transferable attacks are so simple. 
In particular, we found a non-negligible obstacle during our tests: \textit{how should we train a surrogate model}? 
We needed a dataset to train a surrogate model, but we had no clue about the victim's dataset. Furthermore, suppose we were willing to produce a new dataset (or use an external one) for an inherently imbalanced task (e.g., a few samples of a botnet and thousands of benign samples): is the ground truth distribution matching the victim's one? And finally, what is the victim's ML architecture? 
Settings that differ from the victim might negatively impact the attack's success. 
\par
\paragraph{Contributions}
Attackers, therefore, live in a state of ``uncertainty'' when training a surrogate model.
Current literature fails to consider such scenarios, resulting in a lack of understanding of the real effect of state-of-the-art attacks. 
This work fills such a gap by proposing the \textbf{DUMB} attacker model, a framework that allows analyzing if evasion attacks fail to transfer when the
training conditions of surrogate and victim models differ.
In particular, DUMB faces the following conditions:
\textbf{D}ataset so\textbf{U}rces, \textbf{M}odel architecture, and the \textbf{B}alance of the ground truth.
\par
We then propose a novel testbed to analyze the evasion attacks' transferability with DUMB. The testbed consists of three distinct computer vision binary tasks, two sources that generate such datasets, four ground truth balancing levels (from balanced to highly imbalanced), and three models architecture. With this testbed, we analyzed the transferability of seven popular state-of-the-art attacks and six simple image transformations and generated 13K tests.
Such extensive analyses allowed us to unveil new aspects of the transferability of evasion attacks and, furthermore, confirmed the importance of considering the three dimensions introduced with the DUMB attacker model.
\par

Our contributions can be summarized as follows:
\begin{itemize}
    \item We propose the \textbf{DUMB} attacker model, a novel evaluation system to measure evasion transferability. 
    \item We propose a novel testbed to evaluate evasion transferability with the DUMB attacker model. The testbed comprises three distinct computer-vision tasks, four distinct balance levels of the classes, and three distinct state-of-the-art models. 
    \item An extensive evaluation of state-of-the-art evasion attacks with the DUMB attacker model.
\end{itemize}

\paragraph{Findings.}
After evaluating many evasion attacks on all possible combinations of dataset source, model architecture, and class balance of the datasets, our findings can be summarized as follows:
\begin{enumerate}
    \item Less robust models are more susceptible to adversarial perturbations than highly performing models.
    \item Adversarial attacks in literature face difficulty transferring across architectures.
    \item Simple image obfuscation is an effective offensive strategy.
    \item Adversarial attacks struggle when transferring. 
    \item Not all basic surrogate models are ideal for evading attacks.
    \item The discrepancy in class distributions between surrogate and victim datasets can greatly hinder the effectiveness of evasion attacks. Additionally, targeting the minority class seems to be easier than targeting the majority.
    \item Creating surrogate data can negatively impact the effectiveness of transferable attacks.
\end{enumerate}
Our testbed and experiments are open-source and available at the following link: \url{https://github.com/Mhackiori/DUMB}.

\paragraph{Organization}
This paper is organized as follows. 
Section~\ref{sec.preliminaries} summarizes the literature on adversarial machine learning and transferable attacks. Section~\ref{sec.dumb} introduces the DUMB attacker model. 
Section~\ref{sec.methodology} describes the experimental settings. 
Sections~\ref{sec.results} and ~\ref{sec.conclusion} present the results and conclusions of our work, respectively. 

%% file: Sections/02-Preliminaries.tex
\section{Preliminaries}\label{sec.preliminaries}

\paragraph{Adversarial Machine Learning}
Adversarial machine learning (AML) is the discipline that studies how adversaries can exploit machine learning (ML) algorithms to conduct an attack.  
Adversarial attacks can be classified with the following properties~\cite{barreno2006can}: the \textit{influence}, where attackers can actively affect the training procedure (causative attacks), or they simply do not alter the victims' models (exploratory attacks); the \textit{security violation}, where attackers might attempts to alter victims' model's performance (integrity violation), to make victims' model unavailable (availability violation), or to obtain sensitive information (privacy violation); last, the \textit{specificity} of the attack, if the attack targets a specific set of samples (targeted attack) or generic samples (untargeted attacks).
The definition of an attack is further defined by the attackers' knowledge of the victims' system (e.g., training data, model architecture). In particular, we refer to \textit{white-box} attacks when the attacker has (nearly) perfect knowledge about the victim's system, setting the worst-case scenario; on the opposite, we refer to \textit{black-box} attacks when attackers know a little about the target. 

\paragraph{Evasion Attacks}
This work focuses on \textit{evasion attacks}, where attackers aim to modify an input sample to produce a misclassification in the victim's model. 
Malicious samples $x^*$ can be defined as $x^* = x + r$, where $x$ is the original sample, and $r$ is the perturbation. 
The perturbation $r$ can be obtained through the following optimization process:
\begin{equation}\label{eq.perturbation}
    r = \arg \min_z f(x + z) \neq f(x).
\end{equation}
Here, $z$ is the variable being optimized, which represents the perturbation that is added to the original input $x$ to create the perturbed input $x + z$.
Many ML algorithms do not guarantee that the optimization is linear or convex, so we cannot always find a closed-form solution. 
Prior works propose different approaches to estimate such a perturbation; for instance, the Fast Gradient Signed Method (FGSM)~\cite{goodfellow2014explaining}:
\begin{equation}\label{eq.fgsm}
    x^* = x + \varepsilon \cdot sign(\nabla_xJ(\theta, x, y)), 
\end{equation}
where $\varepsilon$ is small to ensure an ``imperceptible'' perturbation, $J$ is a loss function (e.g., cross-entropy), $\theta$ the parameters of the model $f$, and $y$ the ground truth for the given input $x$. 

\paragraph{Transferable Attacks}
A fascinating aspect of adversarial samples is their ability to potentially fool not only the model $f$ used to find the perturbation $r$ for a given sample $x$ but also unknown models $f'$. 
This behavior has a strong repercussion in cyber-security: attackers can therefore leverage their own model $f$ (named substitute or surrogate model) to produce adversarial samples for the victims model. 
Using a substitute model to generate an attack presents many advantages, such as white-box access. 
Papernot et al.~\cite{papernot2016transferability} defined two distinct transferability scenarios by considering the surrogate and victim models. They referred to \textit{intra-techniques transferability} when the two models share the same architecture (e.g., both logistic regression or both Deep Neural Network), or, vice-versa, to \textit{cross-techniques transferability} when the two models have distinct architecture (e.g., one is a logistic regression and the other a Deep Neural Network).   

\paragraph{Adversarial Attacks in Practice}
The literature primarily covers theoretical aspects of threats in machine learning systems. 
Little is known about attacks in practice, where challenges that occur only in real-life might not be considered in controlled environments. 
Therefore, real-life attacks might be utterly different from what is discussed in the literature~\cite{yuan2019stealthy, apruzzese2022real}.
Consequently, industries might perceive as ``innocuous'' threats that are considered technically attractive by the research community and ``serious'' those that are not. 
For instance, consider Perspective, a toxicity detection model deployed by Google: in their recent report~\cite{lees2022new}, the developers tested their model against a simple NLP attack introduced by Gr{\"o}ndahl et al.~\cite{grondahl2018all} that can be deployed by many end-users rather than more complex - and perhaps unrealistic - attacks studied in the literature. 
\par
A few noticeable works proved the feasibility of attacking deployed ML applications: ``All You Need Is Love'', where simple textual perturbations (e.g., typos) endangered toxicity detectors~\cite{grondahl2018all}; 
``stealthy porn'', where researchers showed that social network users evaded porn detectors by applying simple image filters~\cite{yuan2019stealthy}; 
attacks on deployment libraries, where attackers can exploit vulnerabilities of the libraries utilized to deploy a machine learning model~\cite{xiao2018security};
``camouflage attack'', a threat that exploits image-scaling algorithms to produce evasion in computer-vision applications~\cite{xiao2019seeing};
``Zero-Width Space attack'', where invisible Unicode characters inserted in textual samples disrupted the textual representations of many NLP services deployed by top IT companies ~\cite{pajola2021fall};
``Captcha attack'', where researchers showed potential adversarial samples utilized by Instagram users that endanger the OCR of automatic content moderators~\cite{conti2022captcha}. 

\paragraph{Challenges of Transferable Attacks}
Practical constraints might affect the transferability of the attacks as well.
We now summarized relevant prior works that attempted to study different variables that might impact the attacks' transferability. 
Generally, such works are guided by a common observation: it is unrealistic that attackers have knowledge of the victims' systems (e.g., dataset, model architecture), limiting the adoption of surrogate models. 
For instance, training a surrogate model might be expensive (or even impossible) for an attacker since it requires possessing valid training data. We identified two types of solutions in the literature that relax the constraint of having valid data: (i) cross-domain perturbations, i.e. perturbations computed on a task (e.g, paintings, cartoons, or medical images) that transfer on models trained on a distinct task (e.g, ImageNet classes)~\cite{naseer2019cross}; (ii) data-free attacks, where the substitute model can be learned thanks to the cooperation between a generative model, a discriminator, and a series of queries to the victim's model~\cite{zhang2022towards}.
\par
Nevertheless, many works analyzed the impact of surrogates on transferable attacks. 
Mao et al.~\cite{mao2022transfer}, instead, discuss the problem of transferring attacks among computer-vision Machine-Learning-as-a-Service (MLaaS) and analyze how different models' properties might impact the attack. For instance, the authors found that simple surrogates do not necessarily improve transferability and that there is no dominant architecture for surrogates. 
Suciu et al.~\cite{217486} proposed FAIL attacker model, where the authors investigated the impact of evasion transferability under different types of knowledge of victims' systems: the feature space, the architecture of the model, the label instances, and the leverage (i.e., constraints on the type of modification at the feature space).
\par
Compared to the previous works, with \textbf{DUMB}, we attempt to cover unique aspects of the surrogate training, and in particular 
\textbf{D}ataset so\textbf{U}rces, \textbf{M}odel architecture, and the \textbf{B}alance of the ground truth. 
In particular, while aspects like the impact of the model architecture have been covered in literature, others were not, like the source of data and the imbalance problem. Therefore, analyses combining these three aspects are, per se, novel, and they can unveil unique patterns of adversarial transferability. 

%% file: Sections/03-Methodology.tex
\section{The DUMB Attacker Model}\label{sec.dumb}

Suppose being in the shoes of an attacker aiming to evade a victim organization $f'$. 
What are the steps necessary to conduct a (potentially) successful attack? 
Current literature studies the effect of transferability on settings far from being real~\cite{grosse2022so}.
Consider the adversary pipeline necessary to generate an adversarial sample; it consists of: (i) finding a suitable dataset that matches the victim's, (ii) choosing a surrogate model $f$, and picking a methodology that produces adversarial attacks. 
When designing such a pipeline, we find the following challenges that \textit{might} affect the attack execution.
\paragraph{The dataset choice} Prior works mainly use a dataset shared among attackers and victims. \textit{This is unrealistic}. Building a proper surrogate dataset is all but trivial since attackers and victims might follow different corpus generation strategies. 
For instance, in the hate speech detection task, Gr{\"o}ndahl et al. ~\cite{grondahl2018all} show that prior works tackling hate speech propose many datasets following distinct generation procedures; as a result, models trained on a specific dataset lack in terms of generalization on distinct ones. Therefore, in such cases, the transferability might be a property not fully guaranteed. 
    
\paragraph{Ground truth distribution} Prior works mainly assume that attackers and victims use datasets originating from the same source and, therefore, the distributions of the ground-truth match. This is a hard constraint in real settings since such distributions might differ for many reasons. First, the two distributions might result from two distinct methodologies to produce the datasets (see \textit{the choice of the dataset}). Second, many preprocessing techniques might be used to augment the training data. This scenario is likely especially when the task is inherently imbalanced (e.g., hate speech detection\footnote{In the hate-speech detection, usually datasets are strongly imbalanced toward the hateful class~\cite{grondahl2018all}.}). Augmentation techniques can over-sample the minority class (e.g., SMOTE~\cite{chawla2002smote}, Generative Adversarial Networks~\cite{frid2018synthetic}) or undersample the majority one.
    
\paragraph{Model selection} Prior works consider this scenario when analyzing the transferability of distinct adversarial attacks. Indeed, attackers and victims might use one of the many state-of-the-art models or custom ones. For instance, only in computer vision, someone might choose among several models to fine-tune, such as VGG~\cite{simonyan2014very} (and its many versions like VGG16 and VGG19) and ResNet~\cite{he2016deep} (e.g., ResNet18, ResNet50).
\par
Considering such challenges, we can clearly see a need to enhance the study of adversarial transferability in many distinct scenarios and not limit empirical evaluations to a few artificial settings. Thus, experiments focusing on white-box (full access to the victims' model) and black-box (little known about the victims' model) might not be representative of the many shades that might occur in real-life. 
We address such a gap by proposing the \textbf{DUMB} attacker model for transferable samples that present many attack scenario cases. 
\textbf{DUMB} considers \textbf{D}ataset so\textbf{U}rces, \textbf{M}odel architecture, and the \textbf{B}alance of the ground truth, potential factors that might affect the transferability of the attacks.
In Table~\ref{tab:rainbow}, we present eight distinct variations of attacks that can occur in a black-box attack, and in particular, potential mismatches between the source (or surrogate) and target (or victim) models. 
Subscript $a$ and $v$ stand for attacker and victim, respectively.  
We highlight that, in real-life conditions, attackers do not know a priori in which attack scenario they are -- except for the white-box case. 

\input{Tables/rainbow.tex}

%% file: Tables/rainbow.tex
\begin{table*}[!htpb]
    \centering
    \caption{DUMB attacker model of adversarial transferable samples. $DU$ = Dataset soUrce, $M$ = Model architecture, $B$ = balance level.}
    \label{tab:rainbow}
    \begin{tabular}{c|p{2cm}|p{10cm}} \toprule
         \textbf{Case} & \textbf{Condition}  & \textbf{Attack Scenario}\\\midrule
         \rowcolor{gray!15}
         \texttt{C1} &
         \parbox[][45pt][c]{12cm}{$DU_a = DU_v$\\$M_a=M_v$\\$B_a = B_v$}
         &
         \parbox{10cm}{
         The ideal case for an attacker. 
         We identified two potential attack scenarios.
         (i) Attackers legally or illegally gain information about the victims' system.
         (ii) Attackers and victims use the state-of-the-art.
         }
         \\\hline
         \texttt{C2}
         &
         \parbox[][45pt][c]{2cm}{$DU_a = DU_v$\\$M_a=M_v$\\$B_a \neq B_v$}
         &
         \parbox{10cm}{
         Attackers and victims use state-of-the-art datasets and model architecture. However, victims modify the class balance to boost the model's performance. This scenario can occur especially with imbalanced datasets.
         }
         \\ \hline 
        \rowcolor{gray!15}
         \texttt{C3}
         &
         \parbox[][54pt][c]{2cm}{$DU_a = DU_v$\\$M_a \neq M_v$\\$B_a = B_v$}
         &
         \parbox{10cm}{
         Attackers and victims use standard datasets to train their models. However, there is a mismatch in the model architecture. This scenario might occur when state-of-the-art presents many comparable models. Or similarly, the victims choose a specific model based on computational constraints. 
         }
         \\ \hline
         \texttt{C4}
         &
         \parbox[][54pt][c]{2cm}{$DU_a = DU_v$\\$M_a \neq M_v$\\$B_a \neq B_v$}
         &
         \parbox{10cm}{
         Attackers and victims use standard datasets to train their models, while models' architectures differ. Furthermore, victims adopt data augmentation or preprocessing techniques that alter the ground truth distribution (balancing). This scenario can occur especially with imbalanced datasets.
         }
         \\ \hline
         \rowcolor{gray!15}
         \texttt{C5}
         &
         \parbox[][54pt][c]{2cm}{$DU_a \neq DU_v$\\$M_a = M_v$\\$B_a = B_v$}
         &
         \parbox{10cm}{
        Attackers and victims use different datasets to accomplish the same classification task. 
        The ground truth distribution can be equal, especially in inherently balanced tasks. 
        Similarly, models can be equal if they both adopt the state-of-the-art.
         }
         \\ \hline
         \texttt{C6}
         &
         \parbox[][63pt][c]{2cm}{$DU_a \neq DU_v$\\$M_a = M_v$\\$B_a \neq B_v$}
         &
         \parbox{10cm}{
         Attackers and victims use different datasets to accomplish the same classification task. 
         Datasets have different balancing because they are inherently generated in different ways (e.g., see hate speech datasets example) or because the attackers or victims augmented them. Attackers and victims use the same state-of-the-art architecture. 
         }
         \\ \hline
         \rowcolor{gray!15}
         \texttt{C7}
         &
         \parbox[][45pt][c]{2cm}{$DU_a \neq DU_v$\\$M_a \neq M_v$\\$B_a = B_v$}
         &
         \parbox{10cm}{
            Attackers and victims use different datasets to accomplish the same classification task. Datasets ground truth distribution matches. Attackers and victims use different models' architecture.
         }
         \\ \hline
         \texttt{C8}
         &
         \parbox[][45pt][c]{2cm}{$DU_a \neq DU_v$\\$M_a \neq M_v$\\$B_a \neq B_v$}
         &
         \parbox{10cm}{
            The worst-case scenario for an attacker. Attackers do not match the victims' dataset, balancing, and model architecture.  
         }
         \\
         \bottomrule
         \multicolumn{3}{l}{\footnotesize{For simplicity, \texttt{C1} corresponds to the white-box setting, where attackers can access the victims' model, including gradients.}}
    \end{tabular}
\end{table*}

%% file: Sections/04-ExperimentalSettings.tex
\section{Methodology (the DUMB testbed)}\label{sec.methodology}
To simulate the eight specific cases presented in the DUMB table (Table~\ref{tab:rainbow}), we design a testbed that considers distinct sources of datasets, different balance levels, and different model architectures.
In this section, we describe our experimental setup, starting from the data collection phase (Section~\ref{subsec:dataset}), the definition of the balance levels (Section~\ref{subsec:groundtruth}), and the choice of the models (Section~\ref{subsec:models}). Finally, we describe the attacks that we use and their implementation (Section~\ref{subsec:attacks}) and our testing methodology (Section~\ref{subsec:testing}).
Our GitHub repository contains the code and datasets to reproduce our experiments.

\subsection{Dataset Sources (DU-dimension)}
\label{subsec:dataset}
In this work, we focused on the transferability of binary classifiers, which is a common setting in many cybersecurity applications (e.g., spam/non-spam, phishing/non-phishing, hate/non-hate speech). 
We focus on computer-vision tasks since most adversarial attacks literature covers this domain.  
We defined three distinct tasks: Bikes\&Motorbikes, Cats\&Dogs, and Men\&Women.
Given the specific requirements of our testbed, the datasets for each task have been manually collected and validated according to the following steps.
\begin{enumerate}
\item \textit{Data Collection} -- We generate two distinct datasets for each binary task by manually collecting images from two popular search engines: Bing and Google. By creating our own dataset instead of using open-source ones, we can ensure their integrity and have more control over the complexity of the task and possible biases. We collect an average of 14264 images for each dataset.
\item \textit{Duplicate Removal} -- Duplicated images in each dataset are discarded with the \textit{difPy}\footnote{\url{https://github.com/elisemercury/Duplicate-Image-Finder}} library. After this procedure, an average of 254 images are removed from each dataset.
\item \textit{Manual Check} -- Through manual inspection, we ensure that the datasets do not contain erroneous samples (e.g., not coherent with the classes, paintings, sketches, or low quality). 
Although this procedure might reduce any bias of having different data validation strategies between attackers and victims, it allows us to reveal the true effect of having distinct sources that generate (theoretically) the same type of data.
On average, we remove 1854 images from each dataset.
\item \textit{Image Selection} -- We randomly selected, for each dataset, 10000 samples equally split among the two classes.
\item \textit{Image Resizing} -- Using Python Imaging Library (PIL)\footnote{\url{https://pillow.readthedocs.io/en/stable/}}, each image is resized to $300\times300$ and converted to RGB. For the resizing process, we used the \texttt{antialias} option provided by PIL to prevent aliasing artifacts.
\end{enumerate}
For each class in each dataset, we split those 5000 samples into training, validation, and test sets with respective ratios of 70\%, 10\%, and 20\% (i.e., 3500 samples for the training set, 500 samples for the validation set, and 1000 samples for the test set). The images contained in the test set will be used not only to first evaluate the models but also to generate the adversarial samples.

\subsection{Ground Truth Balancing (B-dimension)}\label{subsec:groundtruth}
A second (potentially) critical variable is the different class balance levels between attacker and defender. We simulate different balancing levels in the training sets with the following ratios:
\begin{itemize}
    \item \textit{Balanced} -- 50\% minority class, 50\% majority class.
    \item \textit{Weak Imbalance} -- 40\% minority class, 60\% majority class.
    \item \textit{Medium Imbalance} -- 30\% minority class, 70\% majority class.
    \item \textit{Strong Imbalance} -- 20\% minority class, 80\% majority class.
\end{itemize}
For all our tasks, we choose the first class to be the minority class (i.e., Cats, Men, and Bikes), and this choice will be uniform for all balance levels. The number of class samples for each level of ground truth balancing is shown in Table~\ref{tab:distribution}. To achieve this, we fix the number of samples for the majority class and randomly undersample the minority class accordingly. For instance, to obtain a \textit{strong imbalance} for the Cats\&Dogs task, we keep all the 3500 images of Dogs and randomly select only 875 images of Cats. The validation set and the test set are unaffected by this procedure and contain an equal number of samples between the two classes.

\input{Tables/distribution}

\subsection{Model Architectures (M-dimension)}\label{subsec:models}
\begin{figure}[!htpb]
    \centering
    \includegraphics[width = \linewidth]
    {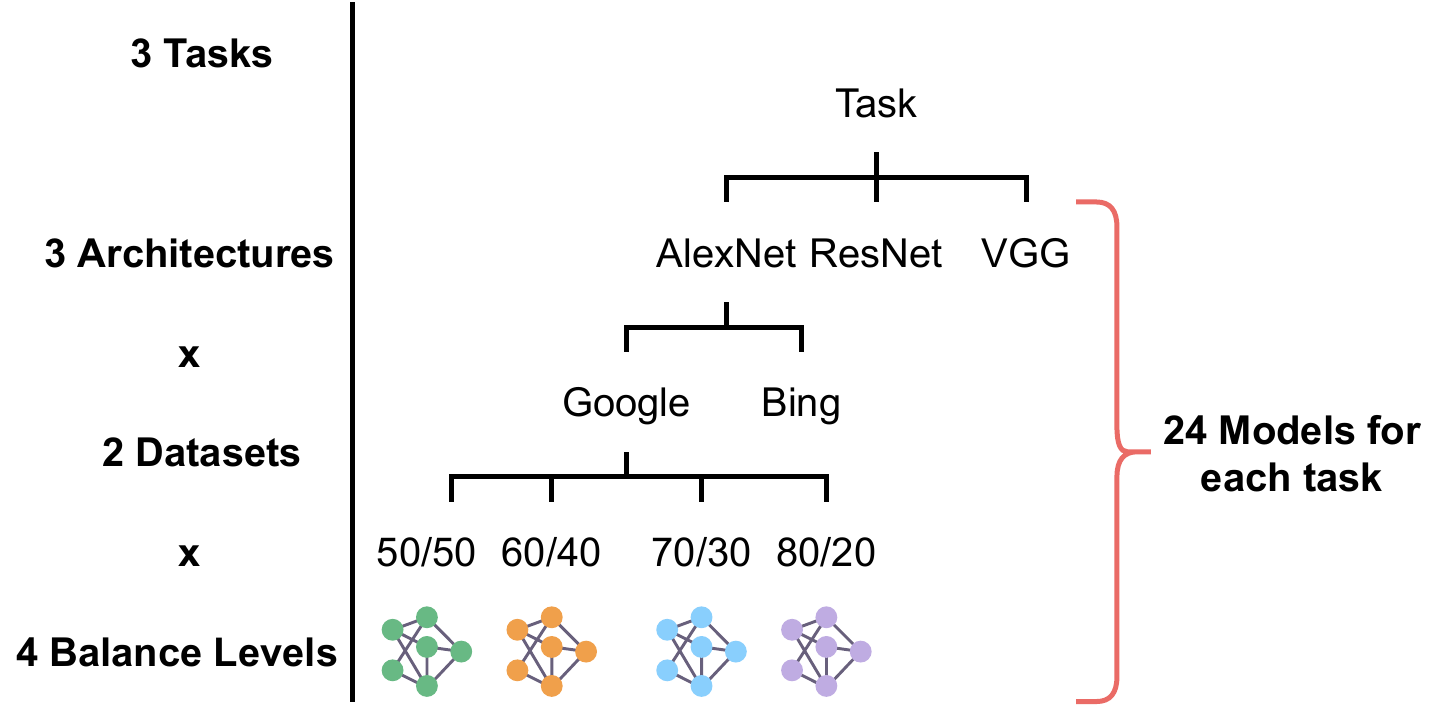}
    \caption{Model combinations during the training phase.}
    \label{fig:modelsCombinations}
\end{figure}

We utilize three state-of-the-art computer vision models for fine-tuning tasks: AlexNet~\cite{krizhevsky2014one}, ResNet~\cite{he2016deep} (ResNet18 version), and VGG~\cite{simonyan2014very} (VGG11-bn version). 
The training procedure follows what is described in the official PyTorch documentation.\footnote{\url{https://pytorch.org/tutorials/intermediate/torchvision_tutorial.html}}
We train a total of 3 tasks $\times$ 2 sources $\times$ 4 class distribution levels $\times$ 3 architectures $=72$ models. A graphical overview of the training combinations is shown in Figure~\ref{fig:modelsCombinations}.

After training the models on each dataset, we evaluate their baseline performance on the test set. As a metric of evaluation, we will use the \textit{F1 score}, which is defined as the harmonic mean of the precision and recall.
This metric provides a balanced measure that considers both aspects of model performance, which is relevant in scenarios with possibly unbalanced dataset distributions.
In particular, the F1 score is expressed as follows:
\begin{equation}\label{eq:f1}
    F1 = 2\frac{precision \cdot recall}{precision + recall}.
\end{equation}
In Table~\ref{tab:baseline-bing} and ~\ref{tab:baseline-google}, we show the average performance of our models at the varying of task, architecture, and class balance levels for models trained on Bing and Google, respectively. 
All models are able to achieve good results on all balancing levels, but some differences can be noticed between the different tasks. Indeed, Men\&Women appear to be the most complex task for any model, while Bikes\&Motorbikes seem to be the easiest among the three.
\input{Tables/baseline.tex}

\subsection{Attacks}
\label{subsec:attacks}
We consider two distinct attack families: \textit{mathematical}, if the result of an optimization process (e.g., FGSM), and \textit{non-mathematical}, if the result of a transformation that does not take into account any machine learning model (e.g., blurring).
\paragraph{Mathematical Attacks}
For the mathematical attacks, we use the following popular attacks.
\begin{itemize}
    \item \textit{BIM} -- Basic Iterative Method adversarial attack, as proposed by Kurakin et al. in their paper~\cite{kurakin2018adversarial}, is a method for generating adversarial examples for image classifiers. The attack works by iteratively perturbing the input image and using gradient descent to optimize the perturbation such that it causes the image classifier to produce the wrong output. One of the key features of the BIM attack is that it can be used to generate adversarial examples that are robust to various types of transformations, such as scaling and rotation.
    \item \textit{DeepFool} -- Moosavi-Dezfooli et al. proposed an algorithm to compute a minimal norm adversarial perturbation for a given image in an iterative manner~\cite{moosavi2016deepfool}. At each iteration, the algorithm adds some perturbation that is computed to take the image to the edge of the region confined by the decision boundaries of the classifier; after that, the perturbations are accumulated to compute the final perturbation, which it is shown to be smaller than the one computed by FGSM in terms of their norm.
    \item \textit{FGSM} -- Fast Gradient Sign Method is one first and simplest adversarial attacks, first proposed by Goodfellow in a paper from 2014~\cite{goodfellow2014explaining}. It works by computing the gradient of the loss of the prediction made by a model based on the true class label of an image and using its sign to construct the adversarial image. 
    \item \textit{PGD} -- Madry et al. proposed the Projected Gradient Descent~\cite{madry2017towards}: an adversarial attack in which an attacker perturbs the input to a machine learning model in such a way as to cause the model to produce the wrong output. The attack works by iteratively calculating the gradient of the loss function with respect to the input and then using this gradient to update the input in the direction that will most likely cause the model to produce the wrong output.
    \item \textit{RFGSM} -- Tramèr et al. proposed an upgraded version of the FGSM attack called Random Fast Gradient Sign Method~\cite{tramer2017ensemble}. The most significant difference is that the FGSM attack generates the perturbation simultaneously, while the RFGSM attack generates the perturbation in a series of "random" steps. This makes the RFGSM attack more computationally efficient, as it can often find an adversarial example faster than the FGSM attack.
    \item \textit{Square} -- Andriushchenko et al.~\cite{andriushchenko2020square} proposed a new black-box attack called Square attack that does not rely on local gradient. It is a score-based attack, meaning that, while not having access to the target model, it can query the probability distribution over the classes predicted by the classifier.
    \item \textit{TIFGSM} -- The paper by Dong et al.~\cite{dong2019evading} proposed a new method for generating adversarial examples, the Translation-Invariant Fast Gradient Sign Method (TI-FGSM), which aims to evade defenses that are based on input transformations by adding a translation-invariant constraint to the iterative FGSM algorithm. The key aspect of the paper is that it achieves high transferability of adversarial examples across different models by making the adversarial perturbations translation-invariant.
\end{itemize}
All mathematical attacks are implemented with \textit{Torchattacks}~\cite{kim2020torchattacks}, a popular python library used in the community~\cite{wu2022dnd, wang2021multi}.
\paragraph{Non-mathematical Attacks}
The other type of attacks we consider is \textit{non-mathematical} attacks. These kinds of attacks do not require any gradient computation and are independent of the model or the task considered. Indeed, non-mathematical attacks have been shown to be effective in real-life ML applications~\cite{yuan2019stealthy}. We implemented these attacks using the PIL library since only simple image processing is required. More in detail, we implemented the following transformations:
\begin{itemize}
    \item \textit{Box Blur} -- By applying this filter, it is possible to blur the image by setting each pixel to the average value of the pixels in a square box extending radius pixels in each direction. It is possible to specify a radius of arbitrary size.
    \item \textit{Gaussian Noise} -- A statistical noise having a probability density function equal to normal distribution. It is possible to specify a $\sigma$ value.
    \item \textit{Grayscale Filter} -- To get a grayscale image, the color information from each RGB channel is removed, leaving only the luminance values. Grayscale images contain only shades of gray and no color because maximum luminance is white and zero luminance is black, so everything in between is a shade of gray.
    \item \textit{Invert Color} -- An image negative is produced by subtracting each pixel from the maximum intensity value, so for color images, colors are replaced by their complementary colors.
    \item \textit{Random Black Box} -- We draw a black square in a random position inside the central portion of the image to cover some crucial information. It is possible to define a size for the black square.
    \item \textit{Salt and Pepper} -- An image can be altered by modifying a certain amount of the pixels in the image either black or white. The effect is similar to sprinkling white and black dots (salt and pepper) in the image. It is possible to specify the proportion of salt and pepper noise.
\end{itemize}
\par
\paragraph{Parameters tuning} All the considered attacks need parameters that regulate the intensity of the perturbations. For instance, all the mathematical attacks have the parameter $\epsilon$, except for DeepFool, which is regulated by the ``overshoot'' parameter. 
Similarly, some non-mathematical attacks have a parameter as well: radius for Box Blur, $\sigma$ for the Gaussian Noise, the size of a black square for the Random Black Box, and the proportion of salt and pepper noise for Salt and Pepper.
In general, we identified optimal parameters $\gamma$ through the following optimization procedure:
\begin{equation} \label{eq.optim}
\begin{split}
& \gamma = \arg \max_s \frac{1}{n}\sum_{i=1}^n f(x_i)\neq f(x_i^*), \\
& \mbox{subject to } \frac{1}{n}\sum_{i=1}^n SSIM(x_i, x_i^*) \geq \alpha.
\end{split}
\end{equation}

\begin{figure*}[!htpb]
    \centering
    \includegraphics[width =\linewidth]{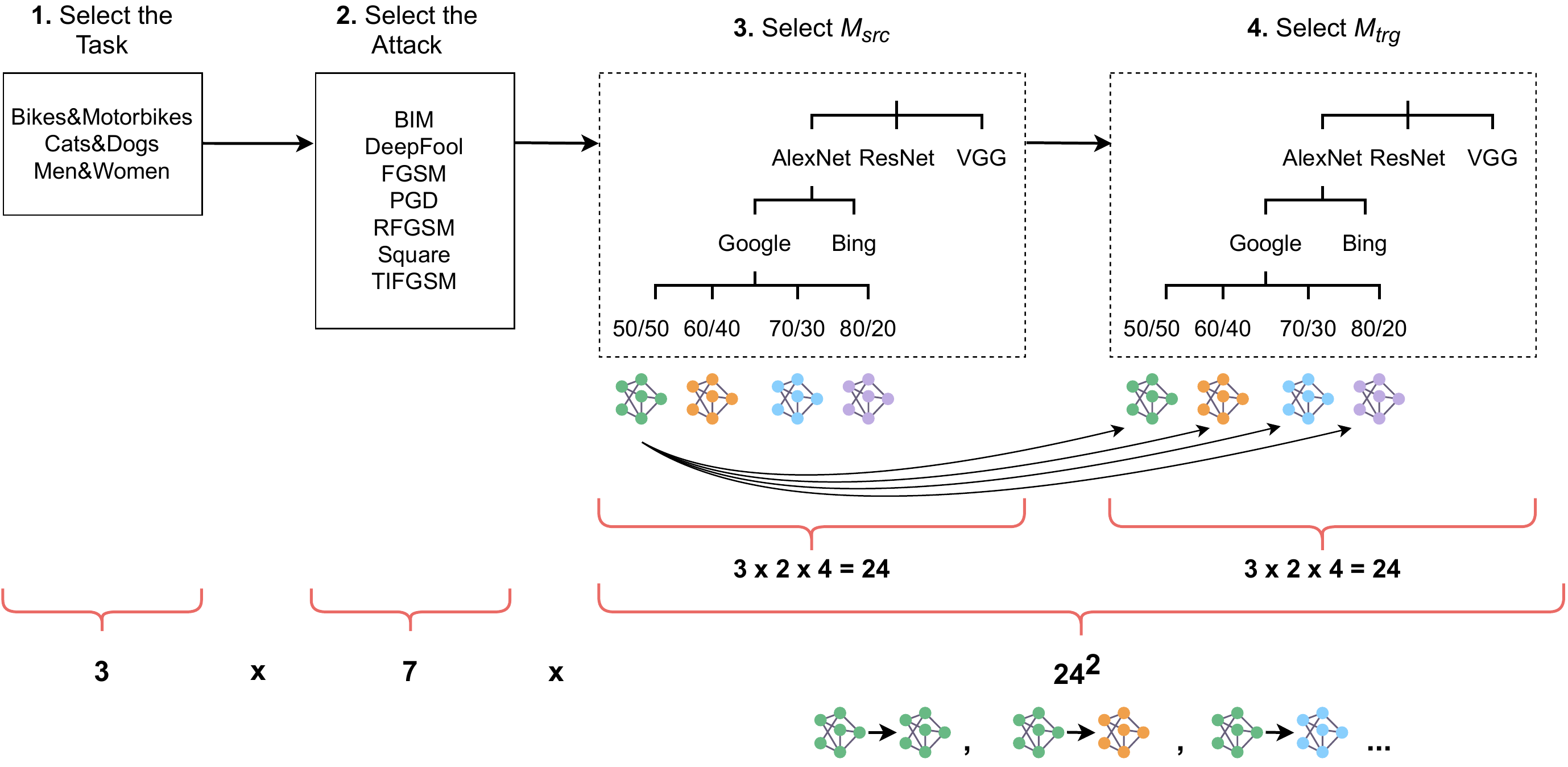}
    \caption{Pipeline of our testing methodology regarding mathematical attacks.}
    \label{fig:testingCombinations}
\end{figure*}

In the notation, $f$ is the model owned by the attacker and used during the optimization process, $x^*$ is the adversarial samples derived by $\mathcal{A}(f, x; s)$, and $\mathcal{A}$ is the adversarial procedure with parameter $s$. 
The reader might notice that the first part of the equation is nothing more than the Attack Success Rate (ASR), where the higher, the more samples evaded. 
The optimization is constrained by the SSIM (Structural Similarity Index Measure), a measure that, given two images, computes their similarity. $\alpha$ is the minimum degradation threshold we accept by the perturbations. In our experiments, we set $\alpha = 0.4$ for all types of attacks. For all mathematical attacks except for DeepFool (i.e., attacks using $\epsilon$ as a parameter), we test $\epsilon$ values in the range $\left[0.01, 0.3\right]$ with a step of $0.01$, while for DeepFool overshoot was tested in the range $\left[10, 100\right]$ with a step of $1$. For non-mathematical attacks with a parameter, ranges and steps were determined individually and based on performance and perturbation. More details on the ranges for the attack parameters can be found in the attack generation script in our repository. 

\subsection{Testing Methodology}\label{subsec:testing}

In Section \ref{subsec:attacks}, we presented a total of 13 attacks, comprising 7 mathematical and 6 non-mathematical attacks. 
After the searching phase for the optimal attacks' configuration (see Equation~\ref{eq.optim}), we generate sets of adversarial samples containing 300 instances equally distributed among the classes.
The images are randomly selected from the corresponding test set, which, however, is filtered in order to consider only images that the model correctly classified. In this way, we ensure that any misclassified adversarial sample can count in the Attack Success Rate.
In the remaining part of the section, we discuss our testing methodology for the adversarial samples against our models separately for mathematical and non-mathematical attacks, as these attacks rely on different approaches.

\paragraph{Mathematical Attacks.} 

Generating adversarial samples for mathematical attacks such as the FGSM requires an input model to compute and generate a perturbed image. Figure~\ref{fig:testingCombinations} shows an overview of the process. For each of the seven attacks we want to test, we need to evaluate all possible combinations of the following pairs $(M_{src},M_{trg})$:
\begin{itemize}
    \item $M_{src}$ -- The model used to generate the adversarial sample. The source model is the surrogate in the transferability setting.
    \item $M_{trg}$ -- The model against which the adversarial sample was tested. The target model is the victim's model in the transferability setting.
\end{itemize}
As explained in Section \ref{subsec:models}, we trained 24 models for each task and used each of them as the $M_{src}$ to generate a set of adversarial samples. We tested each set against 24 different $M_{trg}$, resulting in $24^{2} \times 3$ tasks $ = 1728$ observations for each attack. Since we need to perform this evaluation for each of the seven mathematical attacks, we obtain a total of $1728 \times 7 = 12096$ observations.
\paragraph{Non-mathematical Attacks.}
Non-mathematical attacks, instead, are generated differently since they are transformations applied to the test set of a dataset and do not rely on any model. Thus, for each non-mathematical attack, we generate a total of 2 sets of samples (i.e., the datasets), and we test them on each model, obtaining a total of $2\times24\times3$ tasks $ = 144$. This is valid for each of the 6 non-mathematical attacks we consider, obtaining $144 \times 6 = 864$.

Therefore, the total number of observations performed in our study is $12096 + 864 = 12960$.

%% file: Tables/distribution.tex
\begin{table}[!htpb]
    \centering
    \caption{Number of samples in different levels of imbalance of the training dataset.}
    \label{tab:distribution}
    \begin{tabular}{lll} \toprule
    \textbf{Balance Level} & \textbf{Minority Class} & \textbf{Majority Class} \\ \midrule
    \textit{Balanced} & 3500 & 3500 \\ 
    \textit{Weak Imbalance} & 2334 & 3500 \\ 
    \textit{Medium Imbalance} & 1500 & 3500 \\
    \textit{Strong Imbalance} & 875 & 3500 \\  \bottomrule
    \end{tabular}
\end{table}

%% file: Tables/baseline.tex
\newcolumntype{?}{!{\vrule width 1.25pt}}
\begin{table}[!htpb]
\footnotesize
\centering
\caption{Baseline evaluation of the models. A = AlexNet, R = ResNet, V = VGG.}
\label{tab:baseline}
\begin{subtable}[h]{.5\textwidth}
\centering
\caption{Models trained on Bing.}
\begin{tabular}{p{.6cm}?p{.425cm}|p{.425cm}|p{.425cm}?p{.425cm}|p{.425cm}|p{.425cm}?p{.425cm}|p{.425cm}|p{.425cm}} \toprule
               \parbox[c][12pt][c]{.6cm} & \multicolumn{3}{c?}{\textbf{Bikes\&Motorbikes}} & \multicolumn{3}{c?}{\textbf{Cats\&Dogs}} & \multicolumn{3}{c}{\textbf{Men\&Women}} \\
               \cline{2-10} %
               \parbox[c][12pt][c]{.6cm} & \hfil \textbf{A}    & \hfil \textbf{R}    & \hfil \textbf{V}   & \hfil \textbf{A}  & \hfil \textbf{R} & \hfil \textbf{V} & \hfil \textbf{A}  & \hfil \textbf{R}  & \hfil \textbf{V}  \\ \midrule
\textbf{20/80} & \hfil     0.99        & \hfil  0.99           & \hfil  0.99          & \hfil  0.93         & \hfil   0.97       & \hfil    0.97      & \hfil   0.85     & \hfil   0.92        & \hfil    0.92  \\
\textbf{30/70} & \hfil     0.98        & \hfil  0.99           & \hfil  0.99          & \hfil  0.94         & \hfil   0.97      & \hfil    0.97      & \hfil    0.87       & \hfil 0.93          & \hfil  0.93  \\
\textbf{40/60} & \hfil    0.99        & \hfil   0.99         & \hfil   0.99         & \hfil    0.95       & \hfil    0.98      & \hfil   0.98       & \hfil    0.89       & \hfil 0.94          & \hfil  0.94         \\
\textbf{50/50} & \hfil    0.99         & \hfil    0.99         & \hfil  0.99          & \hfil  0.95         & \hfil  0.98        & \hfil   0.98       & \hfil   0.90        & \hfil 0.95          & \hfil  0.95     \\    \bottomrule
\multicolumn{10}{l}{}
\end{tabular}
\label{tab:baseline-bing}
\end{subtable}
\begin{subtable}[h]{.5\textwidth}
\centering
\caption{Models trained on Google.}
\begin{tabular}{p{.6cm}?p{.425cm}|p{.425cm}|p{.425cm}?p{.425cm}|p{.425cm}|p{.425cm}?p{.425cm}|p{.425cm}|p{.425cm}} \toprule
               \parbox[c][12pt][c]{.6cm} & \multicolumn{3}{c?}{\textbf{Bikes\&Motorbikes}} & \multicolumn{3}{c?}{\textbf{Cats\&Dogs}} & \multicolumn{3}{c}{\textbf{Men\&Women}} \\
               \cline{2-10} %
               \parbox[c][12pt][c]{.6cm} & \hfil \textbf{A}    & \hfil \textbf{R}    & \hfil \textbf{V}   & \hfil \textbf{A}  & \hfil \textbf{R} & \hfil \textbf{V} & \hfil \textbf{A}  & \hfil \textbf{R}  & \hfil \textbf{V}  \\ \midrule
\textbf{20/80} & \hfil  0.96    & \hfil  0.96           & \hfil    0.97        & \hfil   0.95        & \hfil   0.97       & \hfil   0.98       & \hfil   0.82        & \hfil  0.89         & 0.88 \hfil           \\
\textbf{30/70} & \hfil   0.97     & \hfil    0.98         & \hfil  0.98          & \hfil    0.96       & \hfil   0.98      & \hfil   0.98       & \hfil   0.86        & \hfil  0.91         & 0.92 \hfil          \\
\textbf{40/60} & \hfil   0.97    & \hfil     0.98        & \hfil  0.98          & \hfil   0.96        & \hfil   0.98       & \hfil    0.99      & \hfil   0.86        & \hfil  0.92         &  0.93 \hfil           \\
\textbf{50/50} & \hfil  0.97     & \hfil   0.99          & \hfil   0.98         & \hfil  0.96         & \hfil   0.98       & \hfil   0.99       & \hfil   0.86        & \hfil  0.93         & 0.93\hfil      \\    \bottomrule
\end{tabular}
\label{tab:baseline-google}
\end{subtable}
\end{table}

%% file: Sections/05-Results.tex
\section{Results}\label{sec.results}
In this section, we will discuss the evaluation results carried out with our experimental setup.
Given the number of variables that potentially affect our results, we first evaluate the performance of state-of-the-art evasion attacks in the scenarios detailed by the DUMB attacker model (Section~\ref{subsec:dumbperf}). We then evaluate individually the impact of the model (Section~\ref{subsec:modelsimpact}), class distribution (Section~\ref{subsec:classdistributionimpact}), and dataset source (Section~\ref{subsec:sourcesimpact}).
All the raw files from which our results are obtained can be found in the \texttt{results} folder in our repository.

\subsection{DUMB Evaluation}\label{subsec:dumbperf}

\begin{figure*}[!htpb]
    \centering
    \includegraphics[width =.9\linewidth]{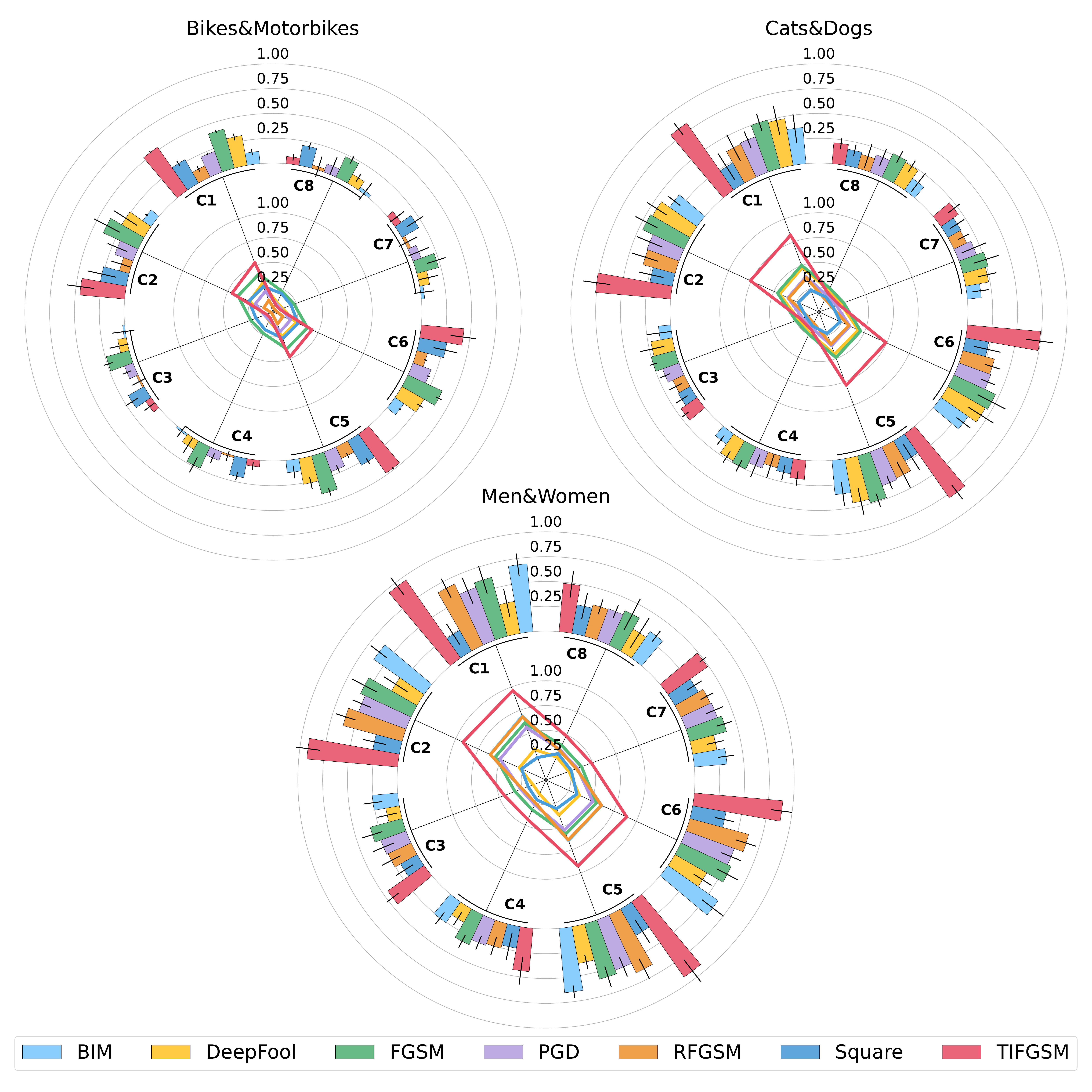}
    \caption{ASR for mathematical attacks. Each subfigure corresponds to a different task and contains two different graphs. The external one shows the performance of our attacks in each of the scenarios of our DUMB attacker model through bar charts. The internal one overviews their overall ASR through a spider chart. While with the former the individual ASR of each attack is more clear, the latter shows their overall performance and trends across the different scenarios. The definitions of the scenarios have been clarified in Table~\ref{tab:rainbow}.}
    \label{fig:overviewMath}
\end{figure*}

In this section, we assess how adversarial attacks perform in the eight distinct cases of our proposed DUMB attacker model. 
We start by analyzing the results of the mathematical attacks, shown in Figure~\ref{fig:overviewMath}. In that Figure, we can observe the effect of two main variables: the task and the attacks. Note that all the performances are averaged among the three DUMB dimensions.

\paragraph{Task}
The first outcome of the analysis highlights how transferability highly varies at the varying of tasks. 
For instance, attacks poorly transfer in Bikes\&Motorbikes, while they are effective in the Men\&Women task.
A possible explanation can be linked with models' performance (reported in Table~\ref{tab:baseline}), where the attack poorly transfers when models greatly solve the task: in the Bikes\&Motorbikes, indeed, models almost perfectly distinguish the two classes, while, on the opposite, on Men\&Women they struggle. 
The outcome suggests malicious actors might easily transfer attacks on models with performances that are far from perfect. 
This finding is concerning if we consider that many real-life tasks are challenging, and state-of-the-art performance is even much below 0.90 of the F1-score. 

\begin{formal}
\textbf{Observation 1:} \textit{Compared with high-performant models, models with performance far from perfect appear more vulnerable to adversarial perturbations.}
\end{formal}

\paragraph{Attacks}
Another noticeable outcome is the superiority of TIFGSM, which outperforms all the other attacks in most cases. We recall that this is the only attack among the considered set explicitly designed for transferability purposes.
The attacks produce a strong transferability in the Men\&Women task, with an evasion rate close to 1 (perfection) in four out of eight cases. 
\par
Last, the ``rectangular'' shape of TIFGSM. By cross-looking with the DUMB attacker model, we can see that TIFGSM, and more in general, all the considered attacks, provide better performance on attacks where attackers and defenders use the same model architecture (i.e., \texttt{C1}, \texttt{C2}, \texttt{C5}, and \texttt{C6}).
Conversely, much lower performance (almost unsuccessful) occurs when attackers and defenders do not share the same model architecture. 

\begin{formal}
\textbf{Observation 2:} \textit{Literature proposes adversarial attacks that struggle to transfer among different architecture.}
\end{formal}

\begin{figure}[!htpb]
    \centering
    \includegraphics[width=0.99\linewidth]{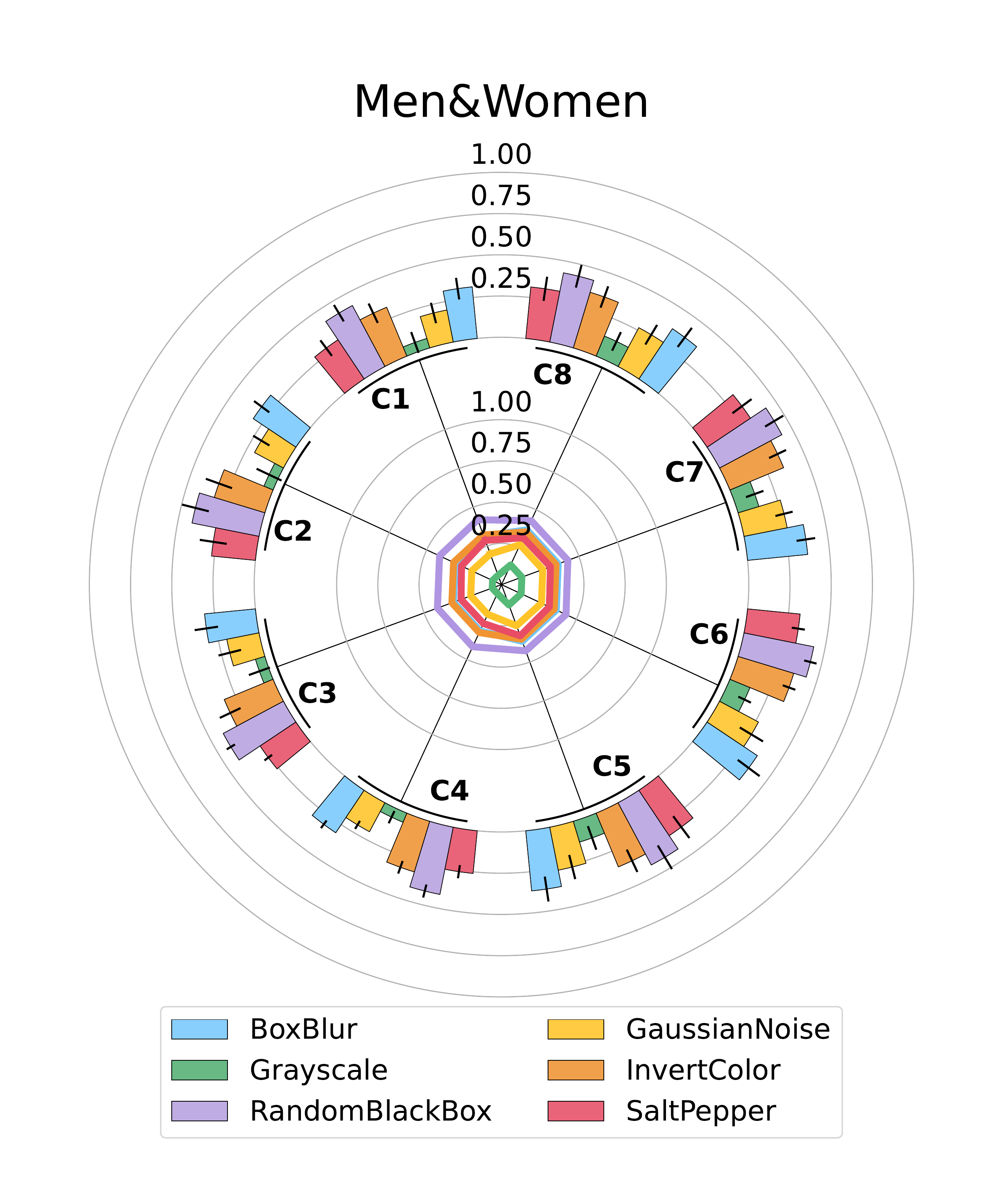}
    \caption{ASR for non-mathematical attacks for the Men\&Women task. Information is conveyed in the same way as Figure~\ref{fig:overviewMath}.}
    \label{fig:overview-nonmath-man}
\end{figure}

\paragraph{Non-Mathematical Attacks}
A different pattern can instead be seen in the non-mathematical attacks, shown to be effective in the past by~\cite{yuan2019stealthy}. For simplicity, we report in the paper only the case of Men\&Women, while more details about the other tasks are available in our GitHub repository. Figure~\ref{fig:overview-nonmath-man} shows the results.
Generally, it appears that simple obfuscations are not effective on our complex models (i.e., AlexNet, ResNet, and VGG). The most effective attack is the RandomBlackBox, which, in contrast, results in the most ``altered'' images.

\begin{figure}[!htpb]
    \centering
    \includegraphics[width=0.99\linewidth]{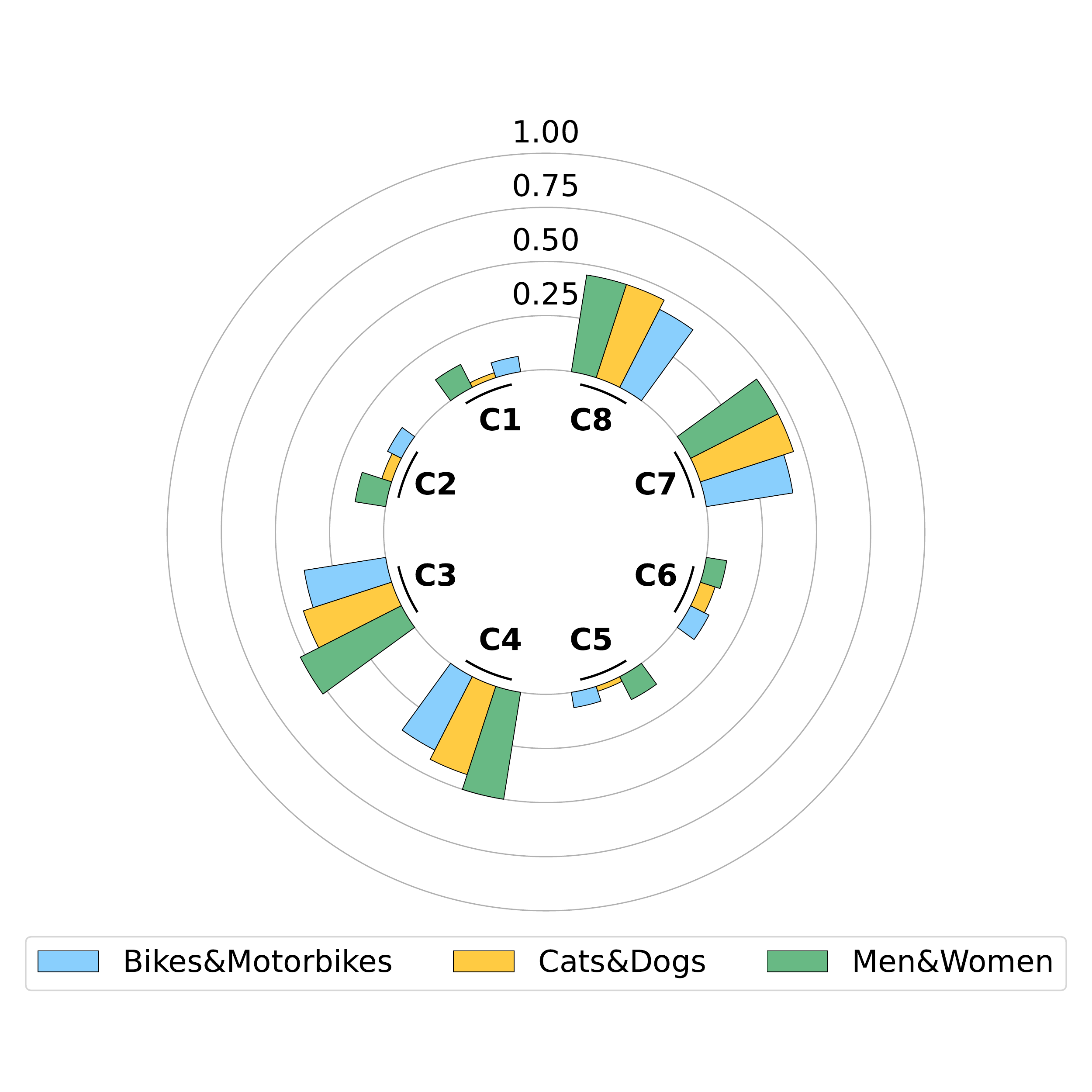}
    \caption{Ratio of non-mathematical outperforming mathematical attacks, grouped by DUMB attacker model cases and tasks.}
    \label{fig:nonmath-vs-math}
\end{figure}

While the TIFGSM generally outperforms non-mathematical attacks, this is not always true for the rest of the considered mathematical attacks.
Therefore, we count how many times non-mathematical outperforms mathematical attacks for each case of the DUMB attacker model and for each task. 
We applied 42 comparisons (7 mathematical $\times$ 6 non-mathematical) for each case, for a total of 336 tests ($42 \times 8$ cases).
Overall, non-mathematical attacks outperform mathematical 79, 81, and 101 times out of 336 cases for Bikes\&Superbikes, Cats\&Dogs, and Men\&Women, respectively. 
Furthermore, we analyzed if such successes are uniformly distributed or centered in some of the DUMB cases. 
The result is shown in Figure~\ref{fig:nonmath-vs-math}.
The reader can observe that the higher values are found in \texttt{C3}, \texttt{C4}, \texttt{C7}, and \texttt{C8}, highlighting the fragility of mathematical attacks in cases where surrogate and victims do not share the same model architecture.

\begin{formal}
\textbf{Observation 3:} \textit{Simple obfuscations are solid offensive black-box techniques.}
\end{formal}

\subsection{Models Impact (M-dimension)}\label{subsec:modelsimpact}
Demontis et al.~\cite{demontis2019adversarial} observed that adversarial transferability depends on the complexity of the surrogate and victim's model. In particular, low-complexity surrogates produce stronger evasion attacks. Similarly, low-complexity victims' models are more resilient to evasions. 
Low-complexity models should be preferred by both attackers and defenders since, for the former, models tend to produce stable gradients that better align with victims' ones. For the latter, models tend to produce smaller gradients size. 

\begin{figure}[!htpb]
    \centering
    \includegraphics[width=\linewidth]{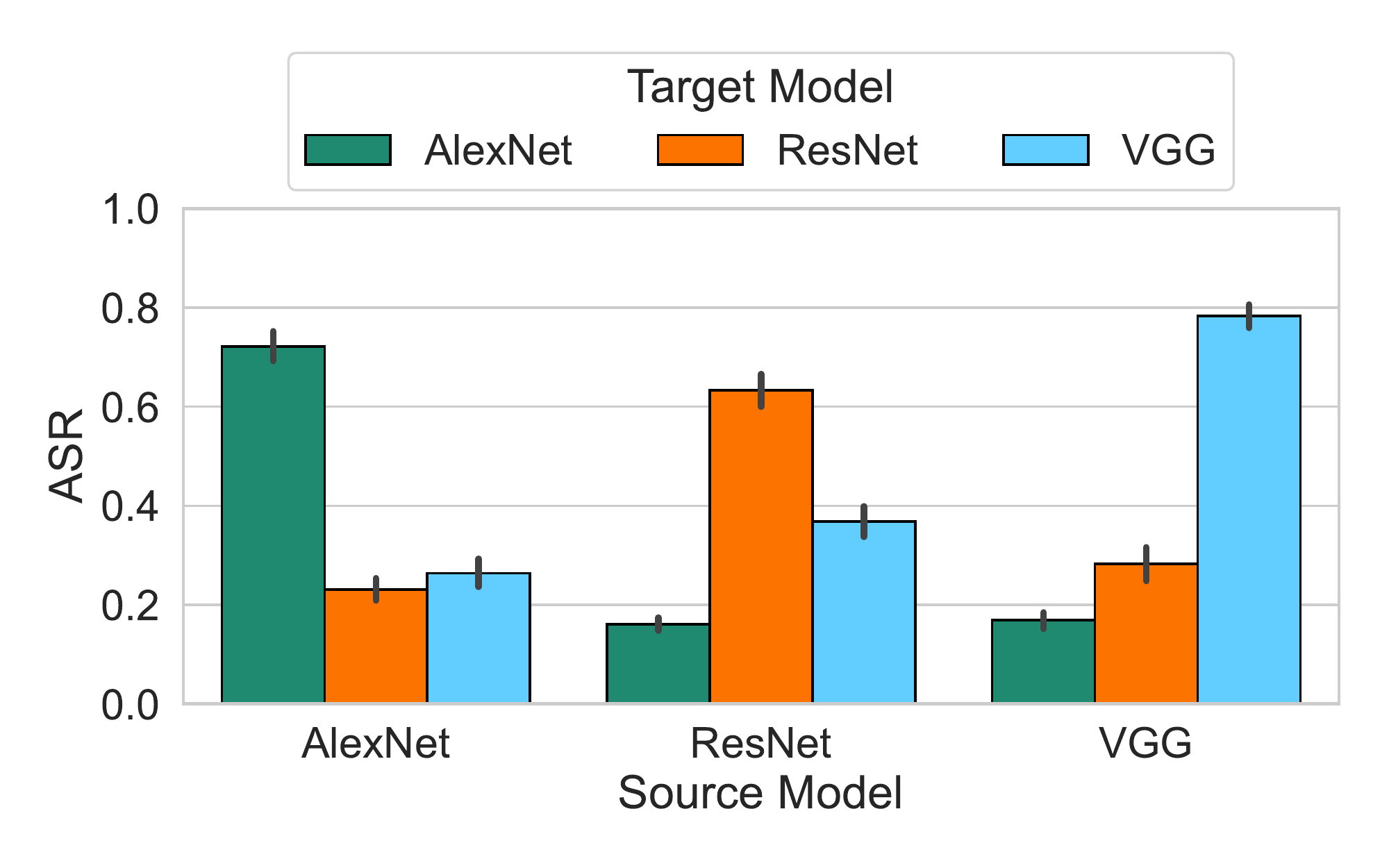}
    \caption{ASR at the varying of source model $M_{src}$ and target model $M_{trg}$. Here, the ``source model'' refers to the model that has been used for adversarial attack generation. In contrast, the ``target model'' refers to the model on which we test the adversarial samples.}
    \label{fig:overview-models}
\end{figure}

Therefore, we now investigate if we observe similar behavior in our testbed. Due to its effectiveness, we focus on the TIFGSM attack. Figure~\ref{fig:overview-models} presents the analysis results by averaging the ASR among the three different datasets. 
We can first observe that, as expected, the highest ASR corresponds to those cases where the source model $M_{src}$ and target model $M_{trg}$ share the same architecture. 
Second, VGG is the weakest victim model for both AlexNet and ResNet. This is shown by the fact that when VGG is the target model, the ASR is the second highest for all other source models (after the case in which $M_{src} = M_{trg}$).
Third, ResNet seems to be the most effective surrogate model. Indeed, when using it as a source model, we see that ASR values are relatively low. At the same time, it is particularly effective as a target model when attacking vulnerable architectures such as VGG.
We find such results not aligned with what was discussed by Demontis et al.~\cite{demontis2019adversarial}, and in agreement with Mao et al.~\cite{mao2022transfer}. 
Consider the complexity of our models, measured in the number of parameters: 61M for AlexNet, 11M for ResNet, and 132M for VGG.  Therefore, ResNet and VGG are the lower and higher complexity models, respectively.

\begin{formal}
\textbf{Observation 4:} \textit{Simple surrogate models are not always optimal to transfer evasion attacks.}
\end{formal}

\subsection{Class Distribution Impact (B-dimension)}\label{subsec:classdistributionimpact}
\begin{figure}[!htpb]
     \centering
     \begin{subfigure}[b]{0.36\textwidth}
         \centering
         \includegraphics[width=\textwidth]{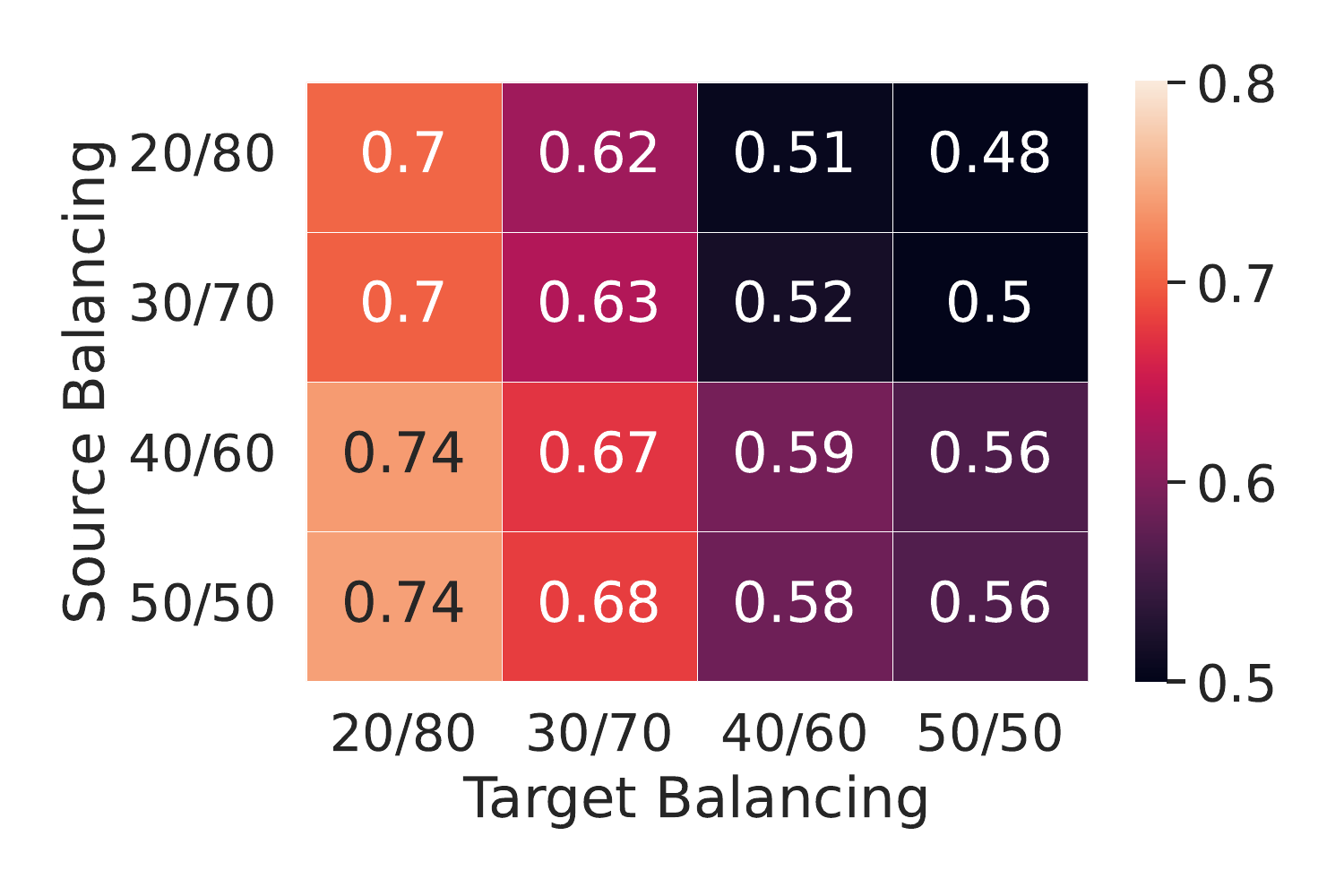}
         \caption{Minority class.}
         \label{fig:min}
     \end{subfigure}
     \begin{subfigure}[b]{0.36\textwidth}
         \centering
         \includegraphics[width=\textwidth]{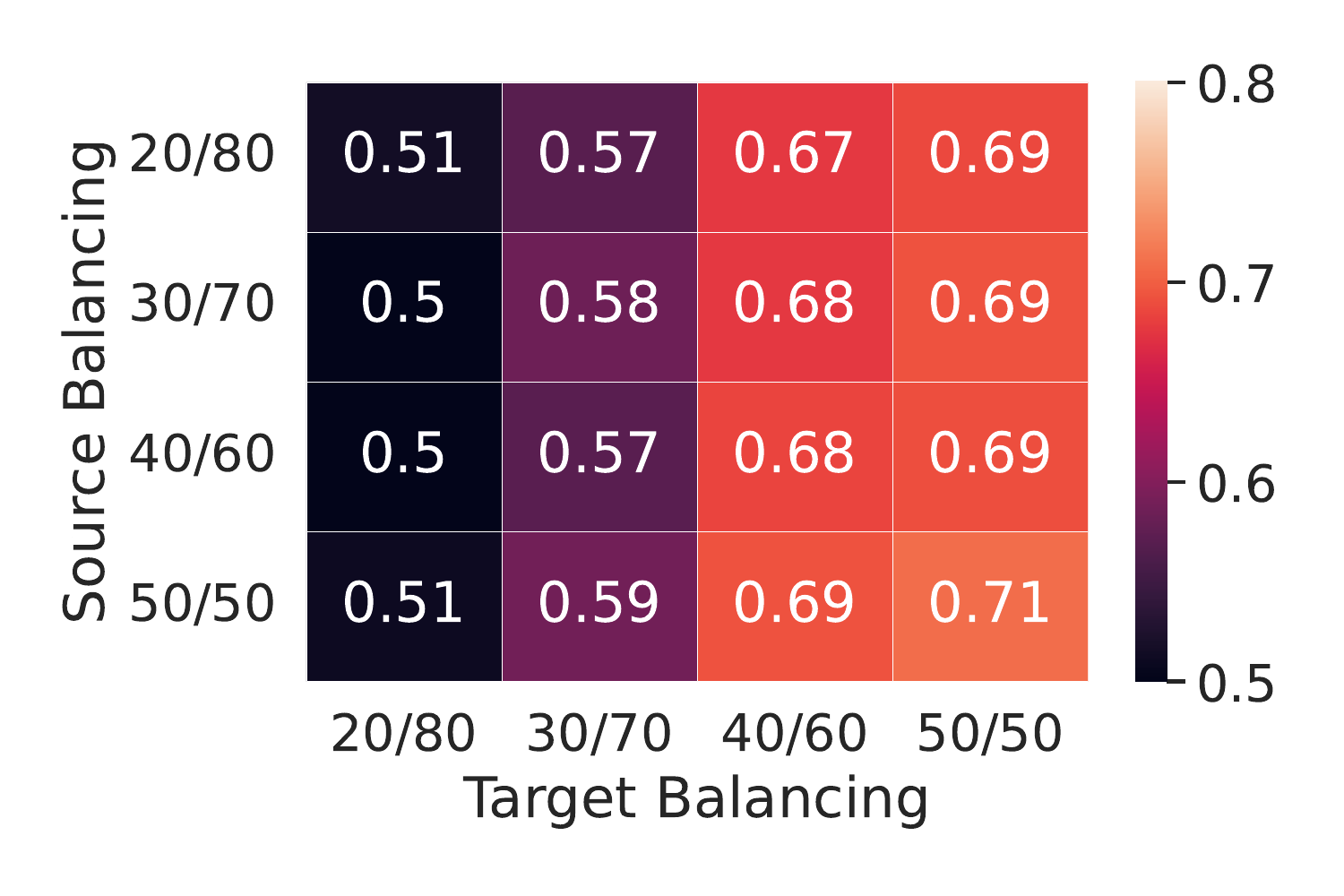}
         \caption{Majority class.}
         \label{fig:maj}
     \end{subfigure}
     \caption{ASR for the minority and majority classes. Here, the ``source balancing'' refers to the balance level of the model that has been used for the adversarial attack generation. In contrast, the ``target balancing'' refers to the balance level of the model on which we test the adversarial samples.}
     \label{fig:heat}
\end{figure}

One of the hypotheses of our work is that attackers and defenders might have different ground-truth distributions. 
Therefore, we investigate how class balance levels impact the success of a transferable attack. 
For simplicity, we show the performance of TIFGSM for the Men\&Women dataset. 
Figure~\ref{fig:heat} shows the results. 
The reader can observe an opposite behavior in the transferability between minority and majority classes. In particular, attacking a minority class under a 20/80 ratio is always effective in every source condition (first column of Figure~\ref{fig:min}). 
The attack increases its complexity as we reach a balancing equilibrium. 
Conversely, it appears to be extremely complex to camouflage a majority sample as a minority one (fist column of Figure~\ref{fig:maj}). 
This observed behavior might be extremely relevant, especially in the context of cybersecurity, where ML classifiers are applied in extremely imbalanced contexts, like malware~\cite{pendlebury2019tesseract} and hate speech detection~\cite{davidson2017automated}, making such applications weak to transferable attacks. 

\par
\begin{figure}[!htpb]
    \centering
    \includegraphics[width=\linewidth]{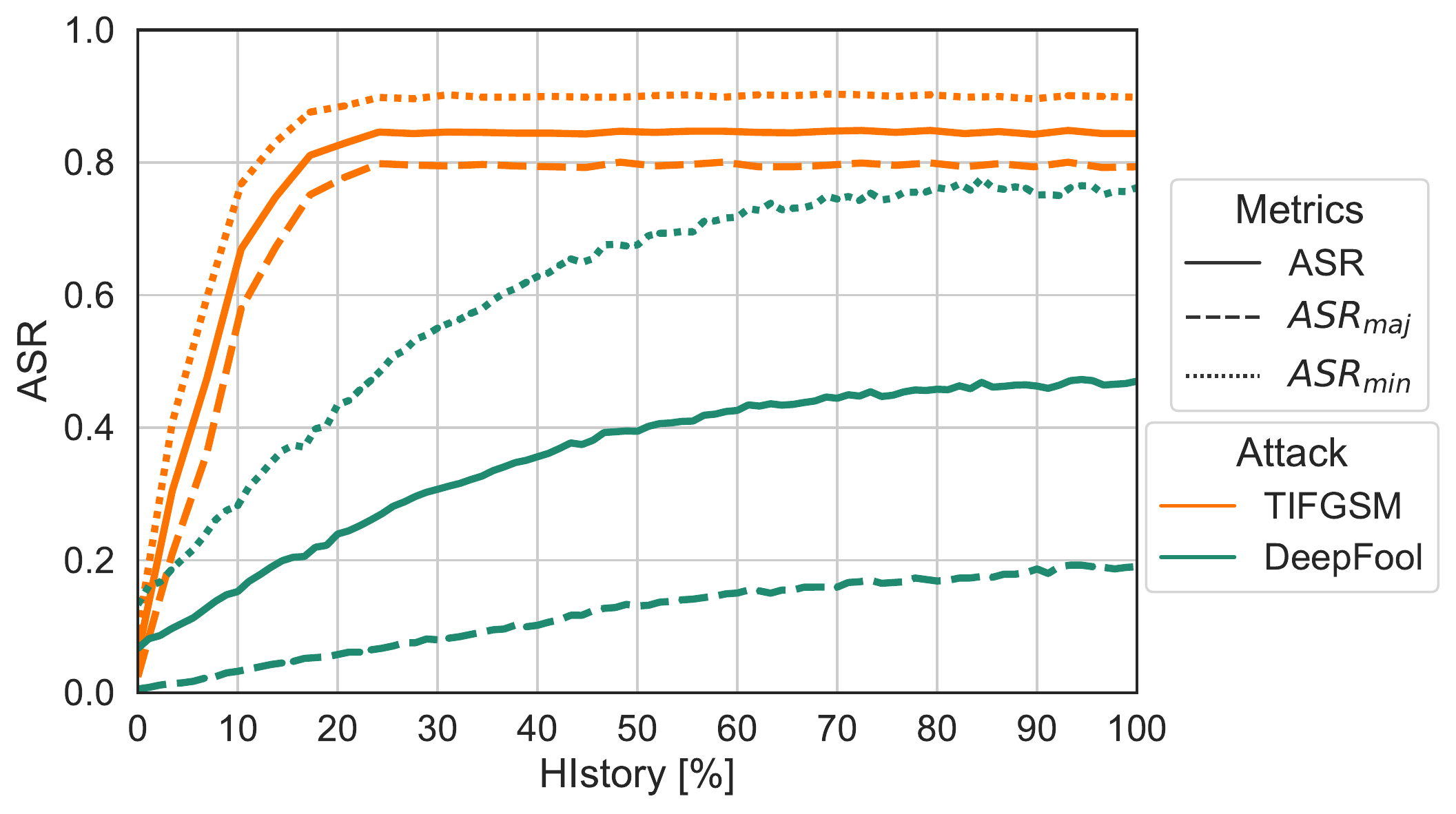}
    \caption{Attacks parameter tuning for Cats\&Dogs dataset, in the 20/80 balance level setting. Since TIFGSM and DeepFool use two different types of parameters with different ranges, we use "history" to characterize their level of perturbation.}
    \label{fig:hist}
\end{figure}

Another important aspect to consider when the source model is trained in an imbalanced dataset is the choice of the perturbation size. 
As we introduced in Equation~\ref{eq.optim}, we computed the global ASR for both classes for each task. However, as shown in Figure~\ref{fig:hist}, the majority class tends to require more perturbation to be effective, while the minority requires a little. 
Therefore, attackers that aim to produce optimal attacks while preserving as much as possible the quality of the samples, should create separate hyperparameter tuning processes for each class.   

\begin{formal}
\textbf{Observation 5:} \textit{The mismatch between the surrogate and victim datasets' class distributions might severely penalize the transferability of the evasion attacks. Furthermore, attacking the minority class appears to be easier compared to the majority.}
\end{formal}

\subsection{Sources Impact (DU-dimension)}\label{subsec:sourcesimpact}
Last, we investigate whether the choice of the dataset impacts the attack transferability. 
The data source has a non-negligible impact if we find at least one case where the choice of the source produces a varied effect on the attack outcome. For example, we can examine the strong imbalance setting (20/80) for the Cats\&Dogs and Men\&Women tasks. This scenario is particularly interesting to study since models typically perform well on the former task but struggle with the latter, often achieving an F1-score lower than 90, as previously shown in Table~\ref{tab:baseline}. We analyze the ASR obtained using the mathematical attacks by considering data sources mismatch, i.e., a surrogate trained on the Bing dataset used to attack models previously trained on the Google dataset, and vice versa. This corresponds to \texttt{C5}, \texttt{C6}, \texttt{C7} and \texttt{C8} of our DUMB attacker model (Table~\ref{tab:rainbow}).

\par
\begin{figure}[!htpb]
    \centering
    \includegraphics[width=\linewidth]{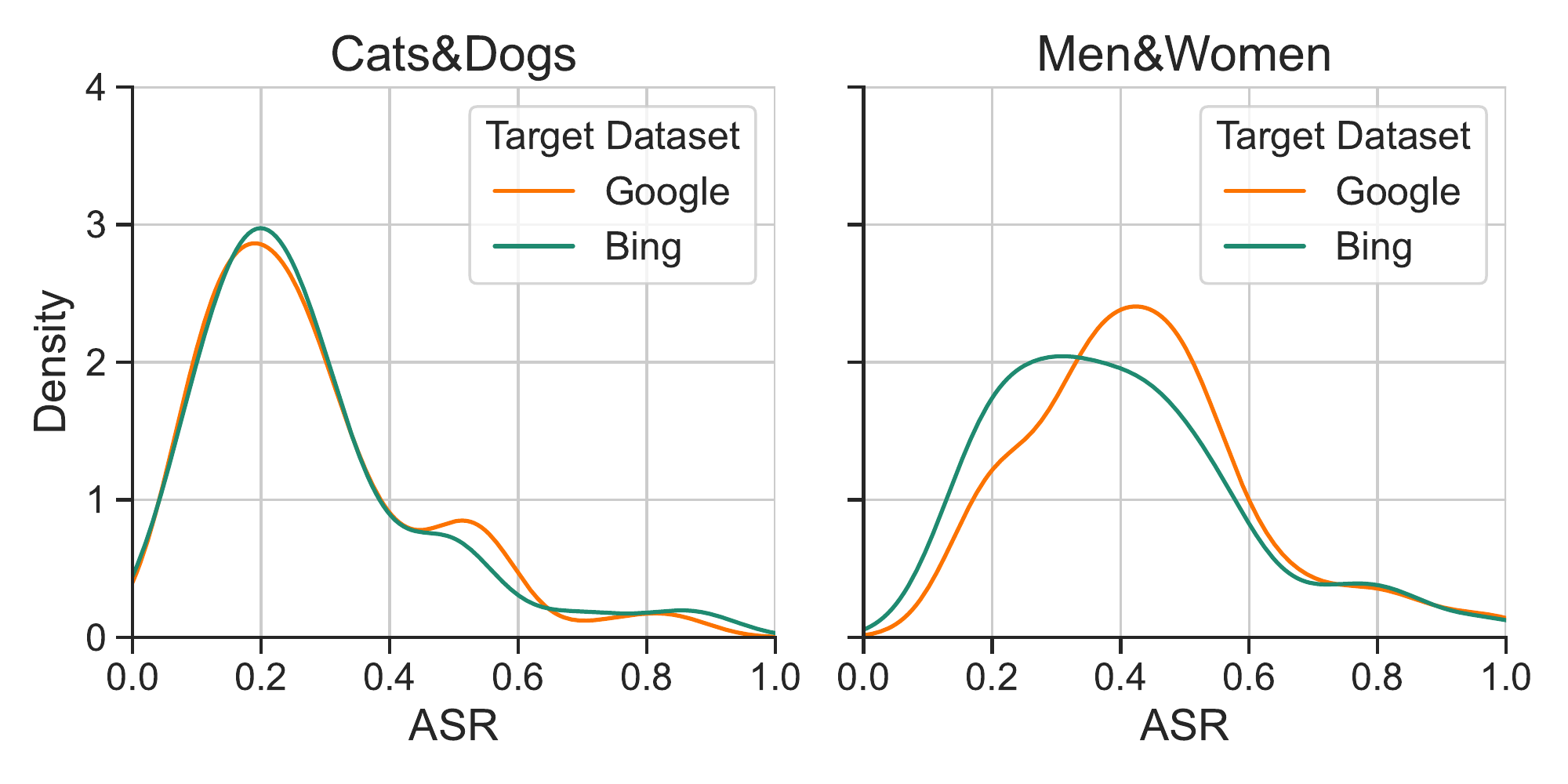}
    \caption{Probability Density Function of ASR for mismatch sources, over 20/80 balance level. 
    The  range of possible ASR is reported on the x-axis, while the y-axis shows the probability density of each ASR. The curve represents the shape of the PDF, where the peak corresponds to the most likely success rate and the width indicates the range of success rates that are probable.
    }
    \label{fig:source}
\end{figure}

Figure~\ref{fig:source} shows the observed distributions. We can notice that for Cats\&Dogs, the two curves almost overlap, while there is a partial mismatch in Men\&Women.
Specifically, regarding the Men\&Women task, it appears that attacks directed toward models trained on the Google dataset (and thus generated from a model trained on the Bing dataset) yield better results. This behavior also reflects the baseline evaluation for the two datasets in Table~\ref{tab:baseline}, where on the same tasks, models trained on Google had lower F1 scores with respect to the ones trained on Bing. 
We statistically confirmed what was observed with the Kolmogorov-Smirnov test (two-sided, the null hypothesis is that the two distributions are equal). We reject the null hypothesis in the Cats\&Dogs case with a $p_{val} = 0.01$.
We, therefore, conclude that the choice of the dataset impacts the attack transferability. 

\begin{formal}
\textbf{Observation 6:} \textit{Generating surrogate data might produce a drop in the performance of transferable attacks.}
\end{formal}

%% file: Sections/06-Conclusion.tex
\section{Conclusion}\label{sec.conclusion}
Transferring evasion attacks among different machine learning models is challenging in a real-world scenario.
While the use of surrogate models has been widely studied in the field of adversarial transferability, many more variables must be considered to depict the full picture of its effectiveness.
\par
In this work, we fill such a gap by proposing the \textbf{DUMB} attacker model. 
This framework allows analyzing if evasion attacks fail to transfer when the training conditions of surrogate and victim models differ.
This framework considers three distinct conditions: \textbf{D}ataset so\textbf{U}rce, \textbf{M}odel architecture, class \textbf{B}alance of the dataset.
Therefore, surrogate and victim models might vary based on the combinations of these conditions, e.g., surrogate and victim models are trained on the same dataset and ground-truth distribution, but they use different architectures. 
\par
We evaluated the DUMB attacker model on our novel \textbf{DUMB} testbed, consisting of 3 distinct binary computer-vision tasks, with two dataset versions each -- collected with Bing and Google as sources --, 4 type of imbalance conditions (from balanced to highly imbalanced), and 3 state-of-the-art model architectures. 
By analyzing 7 well-known evasion attacks and 6 simple image transformations, we explored a total of 13K attacks. 
\par
\paragraph{Considerations}
Our extensive evaluation unveiled aspects that were ignored in the literature or not extensively investigated, with the following repercussions:
\begin{enumerate}
    \item The complexity of the task might have a direct impact on the success of evasion attacks' transferability. As shown in Section~\ref{subsec:dumbperf}, models showing lower performance on the task appear less robust to adversarial attacks. Future works should better investigate the interplay between performance and robustness. 
    \item The above point has a direct impact on real-life machine-learning applications. In particular, often, such tools show performance far from being perfect. This results in tools that are more prone to fail in the presence of adversaries. Therefore, the cybersecurity community should utilize both toy-sh and real-world tasks, where with the former, researchers can gain insights about attacks, and with the latter, adapt such insights to complex scenarios. 
    \item In general, it appears that evasion attacks fail to transfer when the training conditions of surrogate and victim models differ. Future researchers might benefit from both the \textbf{DUMB} attacker model and the \textbf{DUMB testbed} to analyze the transferability of novel proposed attacks. 
    \item While the literature extensively covers the effect of model architecture, little is known about the impact of dataset source and class balancing. For the former, the data generation and labeling process might introduce biases that might impact the transferability. For the latter, many tasks are inherently imbalanced (e.g., spam/non-spam, malware/non-malware), and due to data generation processes or undersampling/oversampling strategies, it is likely that attacker and victim datasets present different ground truth distributions. 
    \item Targeting different classes might lead to different transferable performances. Little attention has been given to the properties of the target class we aim to attack. For instance, when considering the MNIST dataset, the choice seems arbitrary. On the opposite, in cybersecurity tasks, the usual class is the malicious one (e.g., spam, malware). An important property to consider we observed is its numerosity: minority classes of highly imbalanced datasets appear to require limited perturbations to fool (see Figure~\ref{fig:hist}). Future researchers should include such a consideration since many real-life tasks are inherently highly imbalanced, especially those covered by the cybersecurity community. 
    \item We did not observe a model architecture superior in acting as a surrogate model. Future researchers should better investigate the interplay between complex model architectures and their ability to generate transferable attacks.
\end{enumerate}

\paragraph{Limitation and Future Work}
In this study, we aim to provide a systematic view of factors affecting transferability related to the training of a surrogate model. Therefore, some conclusions remain not fully answered and require further studies. 
For instance, our proposed testbed is defined by binary tasks, and our conclusions might not be extended to multiclass tasks. 
Furthermore, our experiment included our novel testbed containing somehow toy-sh tasks, and therefore, far from real conditions. However, our testbed allowed us to clarify different aspects of transferable attacks. Therefore, we believe the proposed testbed might be a precious resource for future researchers conducting analyses in adversarial machine learning.
In particular, we believe that both \textbf{DUMB} attacker model and testbed can be utilized to extend the analyses of attacks, for instance, from evasion to poisoning. 
Moreover, we believe that our work can inspire the definition of novel testbeds, considering, for instance, cybersecurity tasks such as spam and malware detection and network intrusion detection systems.

%% file: Sections/Appendix/A1-Attack_Examples.tex
\section{Additional Experiment Details}
\label{app:details}

\subsection{Attack Examples}

In this paper, we consider many types of attacks to evaluate the different models we train on several datasets and balance levels of the classes. To generate those attacks, as stated in Section~\ref{subsec:attacks}, we must first tune their parameters in order to produce images that maximize the probability of fooling a model while perturbating them as little as possible.

\begin{figure*}[!htpb]
     \centering
     \begin{subfigure}[b]{0.225\textwidth}
         \centering
         \includegraphics[width=\textwidth]{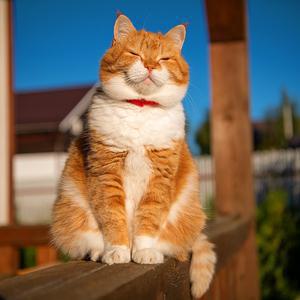}
         \caption{Original.}
         \label{fig:eps0}
     \end{subfigure}
     \begin{subfigure}[b]{0.225\textwidth}
         \centering
         \includegraphics[width=\textwidth]{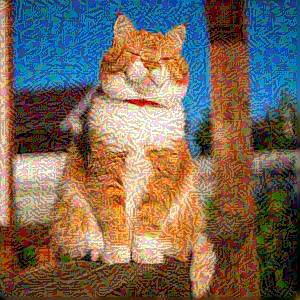}
         \caption{$\epsilon = 0.1$.}
         \label{fig:eps0}
     \end{subfigure}
     \begin{subfigure}[b]{0.225\textwidth}
         \centering
         \includegraphics[width=\textwidth]{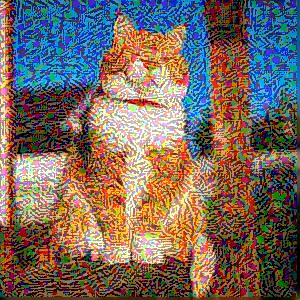}
         \caption{$\epsilon = 0.2$.}
         \label{fig:eps0}
     \end{subfigure}
     \begin{subfigure}[b]{0.225\textwidth}
         \centering
         \includegraphics[width=\textwidth]{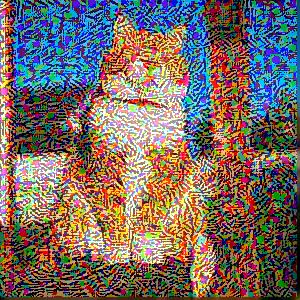}
         \caption{$\epsilon = 0.3$.}
         \label{fig:eps0}
     \end{subfigure}
     \caption{Effect of the attack parameter on the degradation of a sample.}
     \label{fig:perturb}
\end{figure*}

For most of the considered attacks, the level of perturbation added to an image is related to a parameter that can be tuned accordingly. However, by increasing the parameter, the image becomes gradually more degraded until it is unrecognizable even to a human being. For this reason, in Equation~\ref{eq.optim} we included a constraint on the similarity of the image; namely, the SSIM of the perturbed image must be more than $\alpha$. In Figure~\ref{fig:perturb}, we show the increasing degradation of a sample while increasing the attack parameter. More in particular, we show the effect of $\epsilon$ for the FGSM attack on the AlexNet model trained on the Cats\&Dogs task with a strong unbalance level.

\begin{figure*}[!htpb]
     \centering
     \begin{subfigure}[b]{0.225\textwidth}
         \centering
         \includegraphics[width=\textwidth]{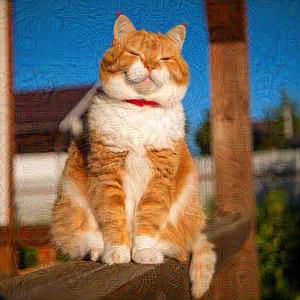}
         \caption{BIM, $\epsilon = 0.29$.}
         \label{fig:app_bim}
     \end{subfigure}
     \begin{subfigure}[b]{0.225\textwidth}
         \centering
         \includegraphics[width=\textwidth]{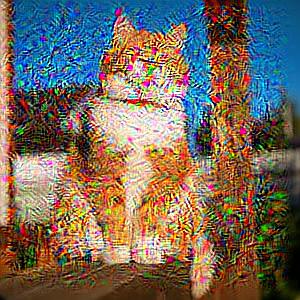}
         \caption{DeepFool, $\mathrm{overshoot} = 0.57$.}
         \label{fig:app_df}
     \end{subfigure}
     \begin{subfigure}[b]{0.225\textwidth}
         \centering
         \includegraphics[width=\textwidth]{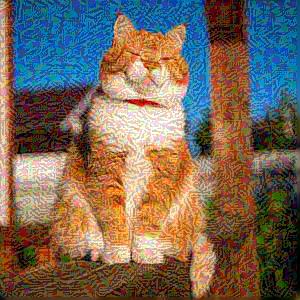}
         \caption{FGSM, $\epsilon = 0.1$.}
         \label{fig:app_fgsm}
     \end{subfigure}
     \begin{subfigure}[b]{0.225\textwidth}
         \centering
         \includegraphics[width=\textwidth]{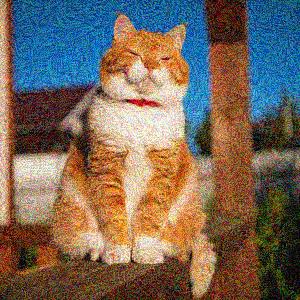}
         \caption{PGD, $\epsilon = 0.22$.}
         \label{fig:app_pgd}
     \end{subfigure}
     \begin{subfigure}[b]{0.225\textwidth}
         \centering
         \includegraphics[width=\textwidth]{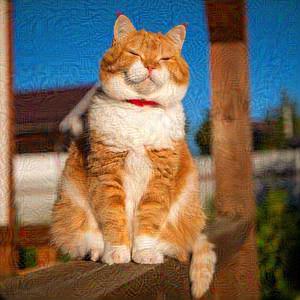}
         \caption{RFGSM, $\epsilon = 0.27$.}
         \label{fig:app_rfgsm}
     \end{subfigure}
     \begin{subfigure}[b]{0.225\textwidth}
         \centering
         \includegraphics[width=\textwidth]{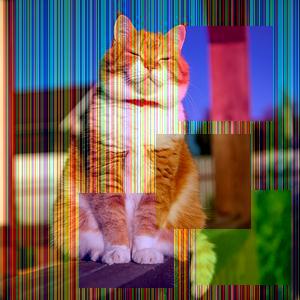}
         \caption{Square, $\epsilon = 0.15$.}
         \label{fig:app_square}
     \end{subfigure}
     \begin{subfigure}[b]{0.225\textwidth}
         \centering
         \includegraphics[width=\textwidth]{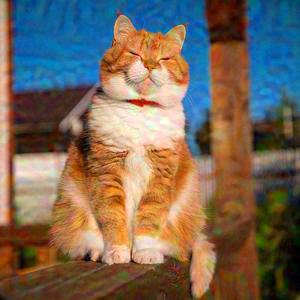}
         \caption{TIFGSM, $\epsilon = 0.28$.}
         \label{fig:app_tifgsm}
     \end{subfigure}
     \caption{Examples of mathematical attacks perturbed with optimal parameter values.}
     \label{fig:appmath}
\end{figure*}

In Figure~\ref{fig:appmath} instead, we extend the visualization to all other mathematical attacks we consider in our testbed. In particular, each sample has been perturbed with its optimal value for $\epsilon$. The samples have been generated to target the AlexNet model trained on the Cats\&Dogs task with a strong unbalance level.

\begin{figure*}[!htpb]
     \centering
     \begin{subfigure}[b]{0.225\textwidth}
         \centering
         \includegraphics[width=\textwidth]{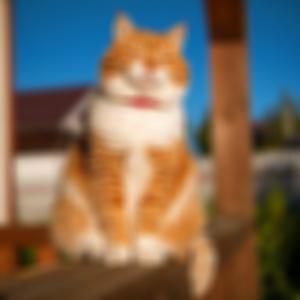}
         \caption{Box Blur, $r = 5.5$.}
         \label{fig:app_bb}
     \end{subfigure}
     \begin{subfigure}[b]{0.225\textwidth}
         \centering
         \includegraphics[width=\textwidth]{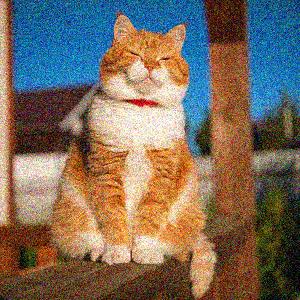}
         \caption{Gaussian Noise, $\sigma = 0.015$.}
         \label{fig:app_gn}
     \end{subfigure}
     \begin{subfigure}[b]{0.225\textwidth}
         \centering
         \includegraphics[width=\textwidth]{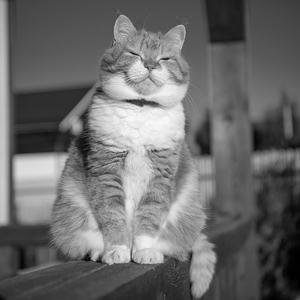}
         \caption{Grayscale Filter.}
         \label{fig:app_gs}
     \end{subfigure}
     \begin{subfigure}[b]{0.225\textwidth}
         \centering
         \includegraphics[width=\textwidth]{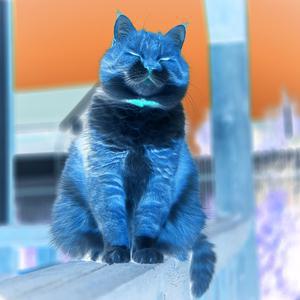}
         \caption{Invert Color.}
         \label{fig:app_ic}
     \end{subfigure}
     \begin{subfigure}[b]{0.225\textwidth}
         \centering
         \includegraphics[width=\textwidth]{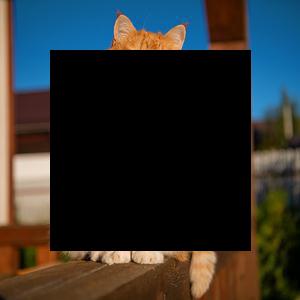}
         \caption{Random Black Box, $\mathrm{size} = 200$.}
         \label{fig:app_rbb}
     \end{subfigure}
     \begin{subfigure}[b]{0.225\textwidth}
         \centering
         \includegraphics[width=\textwidth]{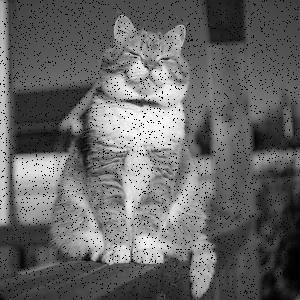}
         \caption{Salt and Pepper, $\mathrm{amount} = 0.05$.}
         \label{fig:app_sap}
     \end{subfigure}
     \caption{Examples of non-mathematical attacks perturbed with optimal parameter values when possible.}
     \label{fig:nonappmath}
\end{figure*}

Finally, in Figure~\ref{fig:nonappmath}, we show the same effect for non-mathematical attacks. Since some of those image transformations also include a parameter, we also include its optimal value.

\subsection{Hardware and Software Configuration}

All experiments have been conducted on a workstation with the following configurations:
\begin{itemize}
    \item \textbf{Operating System}: Ubuntu 20.04.4 LTS.
    \item \textbf{CPU}: AMD Ryzen 5 3600X.
    \item \textbf{GPU}: NVIDIA RTX 3090.
    \item \textbf{Software}: Python 3.8.10, Pytorch 1.7.1.
\end{itemize}